%% file: main.tex
\pdfoutput=1
\UseRawInputEncoding 

\documentclass[]{aastex701}

\usepackage{amsmath}
\usepackage[utf8]{inputenc}
\usepackage[T1]{fontenc}

\usepackage{textcase}
\shorttitle{Search for Neutrinos from Tidal Disruption Events with IceCube}
\shortauthors{IceCube Collaboration}

\begin{document}
\title{Search for Neutrinos from Tidal Disruption Events with IceCube}

\date{\today}

\input{authors}

\begin{abstract}

Tidal disruption events (TDEs) are theorized to produce high-energy neutrinos through photohadronic interactions between accelerated protons and multi-wavelength photons in the accretion disk and outflows. Detecting these neutrinos would provide insight into the dynamics of TDEs. Taking advantage of the recent increase in observed TDEs from wide field-of-view telescopes, we conduct a dedicated search for neutrinos coincident in optical/UV and X-ray wavelengths. We searched for neutrino emission from 89 TDEs selected based on X-ray and optical/UV observations using time-dependent likelihood analysis methods in two parts. First, we searched for emission from individual sources, where we fit the time window of expected neutrino emission. Second, we performed a study of jetted and non-jetted TDE subpopulations using a stacking search with a fixed one year time window. No significant neutrino excess was observed in either search. We set upper limits to the contribution of jetted and non-jetted TDEs detected in optical/UV and X-ray wavelengths to the diffuse astrophysical neutrino flux assuming TDEs are standard candles.

\end{abstract}

\keywords{\uat{Neutrino astronomy}{1100} --- \uat{high-energy astrophysics}{739} --- \uat{Tidal disruption}{1696} }

\section{Introduction}\label{sec:introduction}
Tidal disruption events (TDEs) are a class of transient astrophysical phenomena that occur when stars pass close enough to supermassive black holes (SMBHs) to be ripped apart by tidal forces. TDEs have the potential to accelerate protons to ultra-high energies ($\geq 10^{18}$ eV). They are able to balance the minimum magnetic field required to confine and accelerate cosmic rays with the maximum magnetic field consistent with synchrotron radiation and photo-pion energy loss \citep{farrar_giant_2009}. Theories of neutrino production in TDEs involve the disruption of a star by a SMBH. The stellar disruption provides protons and heavier nuclei which can be accelerated to ultra-high energies within a jet. The accelerated particles interact with a dense photon field, and produce 10\% or more of the observed diffuse high-energy neutrino flux around $1$ PeV \citep{wang_extreme_2012, Dai_Fang_2017, PhysRevD.95.123001, guepin_ultra-high-energy_2018, biehl_tidally_2018}. The mass accretion following the tidal disruption of a star can produce months- or year- long optical/UV and X-ray flares. The neutrino emission could scale with the mass accretion rate resulting in delayed or constant neutrino emission over a timescale of hundreds of days \citep{vanvelzen_establishing_2024}. Other possible scenarios include choked jets in X-ray dark TDEs, where neutrinos are produced in interactions between the jet and surrounding materials \citep{senno_high-energy_2017}. 

To identify neutrinos from TDEs, we rely on data from astrophysical neutrino detectors, such as the IceCube Neutrino Observatory. IceCube is a cubic-kilometer of instrumented ice located at the Amundsen-Scott South Pole Station that detects Cherenkov radiation emitted after neutrinos interact with nucleons in the Antarctic ice \citep{aartsen_icecube_2017}. In 2013, IceCube detected a diffuse flux of high-energy (TeV-PeV) astrophysical neutrinos \citep{icecube_collaboration_evidence_2013}. The first evidence of extragalactic neutrino sources came from a high-energy neutrino associated with a gamma-ray flare from the blazar TXS 0506+056 \citep{aartsen_multimessenger_2018}. This was followed by an archival search of IceCube data for neutrino emission from TXS 0506+056 that found a $3.5\sigma$ neutrino excess from 2013-2014 \citep{IceCube_Collaboration_2018}. IceCube then reported the observation of TeV neutrino emission from the nearby active galactic nucleus (AGN) NGC 1068 \citep{abbasi_evidence_2022}  and the observation of diffuse high-energy neutrino emission from the galactic plane \citep{abbasi_observation_2023}. Beyond these reports, multiple independent IceCube analyses have reported $\sim 3 \sigma$ excesses associated with hard X-ray AGN populations \citep{abbasi_probing_2024, abbasi_search_2022, abbasi_search_2022-1, abbasi_evidence_2025, abbasi_icecube_2025, abbasi_search_2025, icecube_evidence_2025}. \cite{Neronov_Savchenko_Semikoz_2024} using IceCube's public data find 3 sigma evidence for both NGC 4151 and NGC 3079.

Despite these exciting developments, the origin of the majority of the diffuse neutrino flux still remains undiscovered. Thus, we want to probe transient astrophysical objects, one of the primary candidates for production sites of the extragalactic neutrino flux, such as TDEs \citep{murase_high-energy_2019}. Before wide field-of-view telescopes, low TDE observation rates made it difficult to study these transients. The Zwicky Transient Facility (ZTF), with a field-of-view of $\sim$ 47 sq degrees, has increased the number of TDE observations. Furthermore, ZTF implemented a systematic neutrino follow-up program that reported the TDE AT2019dsg and accretion flares AT2019fdr and AT2019aalc as probable counterparts to IceCube neutrino alerts \citep{robert_stein_tidal_2020, vanvelzen_establishing_2024}. The neutrino reported by these publications to be coincident with AT2019dsg arrived with an approximately 150 day delay from the optical peak of the flare, which led to theoretical models of TDE neutrino production involving sustained mass accretion or mildly relativistic outflows \citep{piran_disk_2015, roth_radiative_2020, vanvelzen_establishing_2024}. Although a recent release of IceCube public data that used an improved reconstruction technique found the positions of these flares to be well outside the updated containment region of the corresponding neutrinos \citep{icecube_icecat-2_2025}, these associations and related theoretical advances have motivated the search for neutrino production in TDEs. More recently, transients AT2021lwx and AT2022sxl were also reported to be in possible coincidence with IceCube neutrino events \citep{yuan__at2021lwx_2024, ji_at2022sxl_2025}, but these sources were not included in this search because they are not classified as TDEs. A previous IceCube analysis of 3 jetted and 13 unambiguous TDEs limited non-jetted and jetted TDE contributions to the diffuse high-energy neutrino flux to 26\% and 1.3\%, respectively \citep{robert_stein_search_2019}. 

In this work, we present both a single source and stacking analysis on neutrino emission from TDEs using an updated catalog, including a large increase in classifications from wide field-of-view telescopes such as ZTF \citep{velzen_seventeen_2021, hammerstein_final_2022}. Section~\ref{sec:catalog and data} describes the TDE catalog selection as well as the neutrino data used. The methods and results are presented in Section~\ref{sec:sss} for the catalog search and in Section~\ref{sec:stacking} for the stacking search. Section~\ref{sec:realtime} discussion of the reported TDEs coincident with IceCube alert neutrinos. A summary of our findings and future outlooks are presented in Section~\ref{sec:conclusions}.     

\begin{figure*}[ht!]
\centering
\includegraphics[width=0.85\linewidth]{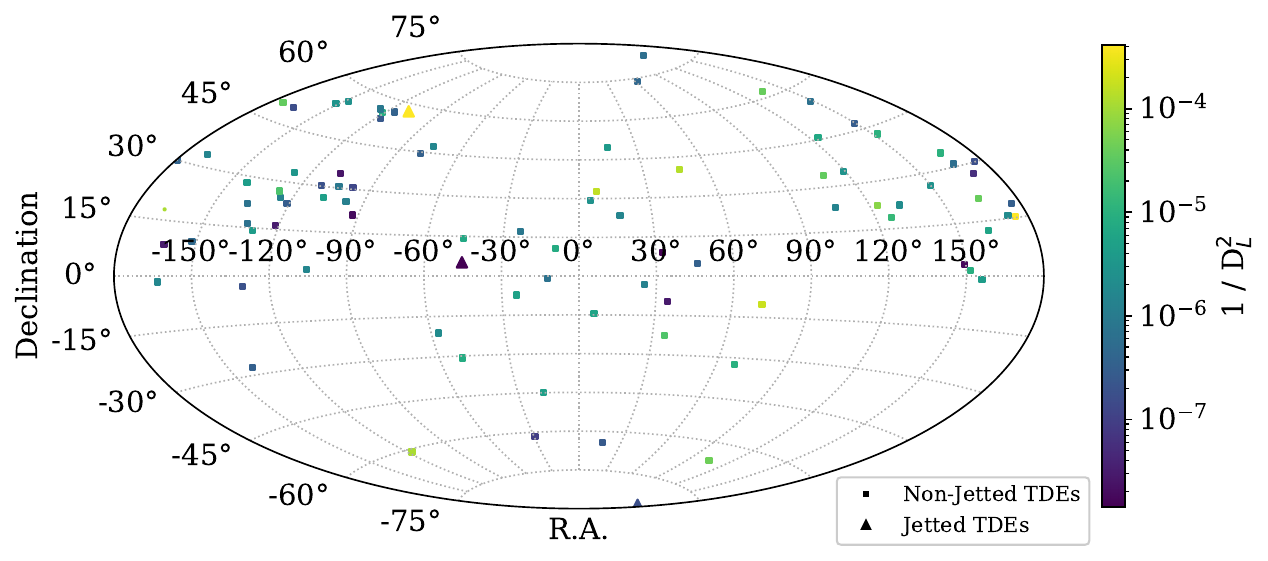}
\caption{\textbf{TDE Catalog.} Source distribution of UV/Optical and X-ray detected TDEs using Equatorial coordinates. The colorbar indicates the inverse square of the luminosity distance, $D_L$, of each source, which is used as the weighting scheme of the stacking search. The catalog consists of 3 on-axis jetted TDEs (triangles) and 86 non-jetted TDEs (squares). 
\label{fig:catalog}}
\end{figure*}

\section{TDE Catalog \& IceCube Data Selection}\label{sec:catalog and data}
Observationally, the mass accretion that occurs during the tidal disruption of a star appears as a months- or years- long flare, with emission over a wide range of wavelengths. In this selection, TDEs with optical detections feature a light curve that follows a power-law decline from peak as expected from the theoretical stellar debris fallback rate  \citep{gezari_tidal_2021}. TDEs selected in soft X-rays feature a spectrum well-fitted by blackbody temperatures ranging from 0.02-0.13 keV that has dramatic variability over time \citep{gezari_tidal_2021}. While these TDEs have different selection methods, both are representative of the overall unified TDE population \citep{guolo_systematic_2024}. The TDE catalog used in this work, shown in Figure~\ref{fig:catalog}, is based on classifications reported in the Transient Name Server (TNS\footnote{\url{https://www.wis-tns.org/}}) and the literature and is therefore subject to selection effects and incompleteness. In particular, the sample is biased toward events with sufficient optical/UV and/or X-ray brightness and follow-up to enable classification. 

From this catalog, we considered TDEs detected in optical/UV and X-rays from 2008 to 2022, including 3 on-axis jetted TDEs and 86 non-jetted TDEs. TDEs identified through optical/UV and X-ray observations are the focus of this study because TDE emission is well-characterized in these wavelengths and can theoretically be associated with the photo-hadronic interactions responsible for neutrino production \citep{winter_interpretation_2023}. This catalog includes 34 TDEs with both X-ray and optical observations, 52 detected optically-only, and 3 only in X-ray. A complete catalog with source information can be found in Table~\ref{tab:catalog} of Appendix~\ref{a:sourcecat}. In Section~\ref{sec:stacking}, this catalog is divided into jetted and non-jetted subpopulations based on observations of bright, rapidly variable X-ray and radio emissions. 

To search for neutrinos coincident with these TDEs, the analyses use neutrino data assembled from \textit{track} and \textit{cascade} events detected by IceCube. \textit{Track} events are predominantly produced by the outgoing muon from charged-current muon neutrino interactions. Tracks are generally preferred for neutrino point source searches because they have small angular errors (< $1^\circ$ at TeV energies) and a high detection rate of $\sim 2$ mHz, providing a robust sample \citep{aartsen_time-integrated_2020}. The track dataset used in this analysis totals about 1.8 million events in about 15 years of exposure (PS Tracks; \cite{Abbasi_datarelease_2026}). \textit{Cascade} events are produced by charged-current electron neutrino and tau neutrino interactions as well as neutral-current interactions of all flavors \citep{aartsen_search_2017}. Cascades, which have been used to find evidence of neutrino emission from the galactic plane \citep{abbasi_observation_2023}, have larger directional uncertainties ($\sim 10^\circ$ at TeV energies), but have a more accurate neutrino energy reconstruction because they are mostly contained within the detector \citep{aartsen_constraints_2017}. The cascade dataset (DNNCascade) used in this analysis adds approximately two years to the dataset used in \cite{abbasi_observation_2023}, totaling about 75,000 events in about 12 years of exposure. The two different detection channels can be simultaneously fit in a maximum likelihood analysis to fully exploit the benefits of both track and cascade event types \citep{abbasi_all-sky_2025}.

\section{Single source catalog search}\label{sec:sss}

\subsection{Analysis methodology}\label{sec:ss_methods}

This analysis utilizes an unbinned extended maximum likelihood method \citep{Braun_2008, Braun_2010}. The likelihood consists of spatial, temporal, and energy probability density functions (PDFs), which characterize how signal-like or background-like a given event candidate is. The null hypothesis assumes that all events come from atmospheric neutrinos and muons or the diffuse all-sky background. The signal hypothesis assumes an additional accumulation of astrophysical neutrinos clustered in space and time around an astrophysical source. The signal energy PDF assumes an $E_\nu^{-\gamma}$ power-law, where the spectral index $\gamma$ is a fit parameter in the likelihood. The signal space PDF uses either a 2D Gaussian or Kent distribution, depending on the angular uncertainty of the evaluated event. The signal time PDF assumes a uniform box probability starting at the UV or X-ray peak with an end time that is fit within the range of 1 second to 1 year post-peak. 

The general form of the likelihood used in this analysis is 
\begin{equation}
    \mathcal{L}(n_s, \gamma, t_1 | \boldsymbol{x}) =\frac{(n_s + n_b)^{N}}{N!}e^{-(n_s +n_b)} \prod_{i}^{N \in \Delta t} \left[ \frac{n_s}{N} \mathcal{S}(\boldsymbol{x}_i | \gamma, \boldsymbol{d}_{src}) + \left(1- \frac{n_s}{N}\right) \mathcal{B}(\boldsymbol{x}_i) \right],
\end{equation}\label{llh}where $\boldsymbol{x}_i$ is a vector consisting of the observed energy, reconstructed direction, and angular uncertainty of the direction for a given event $i$, $\boldsymbol{d}_{src}$ is the position of the source TDE, $N$ is the total number of events, $n_b$ is the number of background events, and $\mathcal{S}$ and $\mathcal{B}$ are the signal and background PDFs, respectively. The free parameters of the likelihood are the mean number of signal events, $n_s$, the spectral index of a fitted unbroken power-law spectrum, $\gamma$, and the time window length of neutrino emission, $\Delta t = t_1 - t_0 $, where $t_0$ is fixed to the UV or X-ray peak and $t_1$ is fit, sampling values from 1 second to 1 year. To evaluate the overall likelihood, we multiply Eq.~\ref{llh} over each dataset used \citep{aartsen_search_2013}.

The test statistic (TS), Eq.~\ref{ts}, is evaluated as the negative logarithm of the likelihood ratio between the background and signal hypotheses, with the signal likelihood evaluated with best fit parameters $\hat{n}_s$, $\hat{\gamma}$, and $\hat{\Delta t}$:
\begin{equation}\label{ts}
    TS \equiv -2 \log \left( \frac{\mathcal{L}(n_s = 0)}{\mathcal{L}(\hat{n}_s, \hat{\gamma}, \hat{\Delta t})} \right).
\end{equation}
Events in temporal and close spatial coincidence with a TDE will result in a large TS value. To make a statement on how consistent the TS value i with the null hypothesis, we calculate a $p$-value. We perform pseudo-experiments, in which we simulate the background TS distribution using a data-based model. Taking advantage of the detector azimuthal symmetry, the data is scrambled by randomly assigning new values of right ascension and time to events. The background TS distributions are fit to a $\chi^2$ distribution that is used to evaluate the significance of the analysis.

In this single source catalog search, one hypothesis is tested for each TDE in the catalog. Given each independent test has a probability to randomly produce an over-fluctuation, smaller $p$-values become increasingly likely as more tests are added, referred to as the look elsewhere effect \citep{ranucci_profile_2012}. To account for this effect, we implement a trials correction by multiplying the $p$-values by the number of sources, which quantifies the number of independent tests performed.

\begin{figure*}[ht!]
\centering
\includegraphics[width=0.85\linewidth]{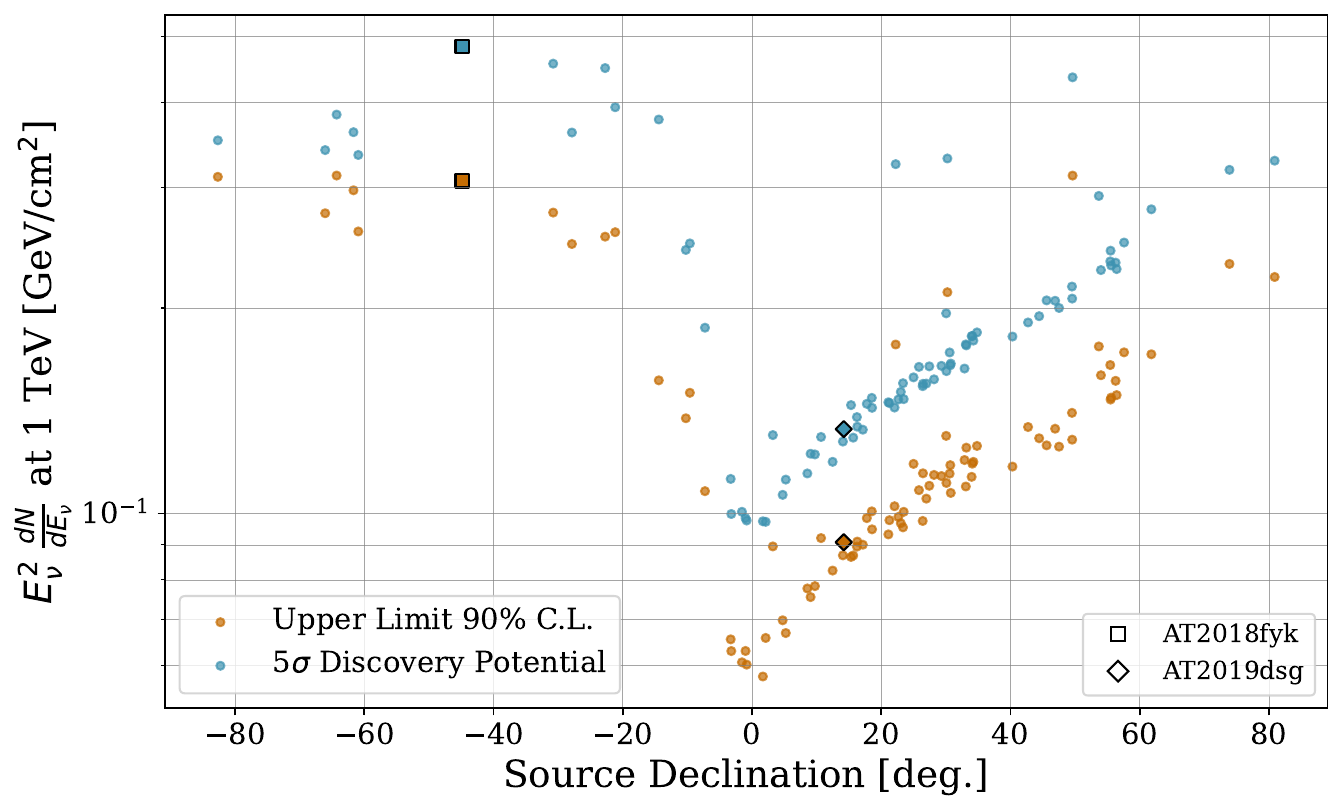}
\caption{\textbf{Per source upper limits.} The 90\% Confidence Level (C.L.) upper limits and median $5\sigma$ discovery potential (D.P.) of the time-integrated flux per TDE are shown, assuming an E$_\nu^{-2}$ spectrum and an expected neutrino emission window of 150 days and searching a maximum time window length of 1 year. $E_\nu^2dN/dE_\nu$ is the per-flavor time-integrated flux, with number of neutrinos ($N$) and neutrino energy ($E_\nu$), assuming a flavor ratio $(\nu_e:\nu_\mu:\nu_\tau )$ of 1:1:1. The locations of AT2019dsg and AT2018fyk are marked with a black square and black diamond outline, respectively, for reference.
\label{fig:ULs}}
\end{figure*}

\subsection{Catalog Search Results}

We observed no statistically significant emission in this catalog search. The most significant identified source was AT2018fyk with a $p$-value of 0.0087 and $1.06\sigma$ significance after trials correction. The best fit values for this source were $n_s = 4.72$, $\gamma = 2.14$, and $\Delta t = 9.13$ days. It is worth noting that AT2018fyk is a particularly interesting TDE that has exhibited repeated brightening episodes following its initial flare \citep{Wevers_2023}. While this does not provide the basis for any claim in this work, re-brightening TDEs with harder X-ray emission could be an interesting source class for a future analysis. Upper limits are calculated at the 90\% confidence level for the time-integrated flux at 1 TeV per TDE and plotted as a function of neutrino energy in Figure~\ref{fig:ULs}. The upper limits assume a power law spectrum with $\gamma = 2$, a 150-day time window corresponding with theoretical neutrino emission timescales \citep{vanvelzen_establishing_2024}, and a neutrino flavor ratio $(\nu_e:\nu_\mu:\nu_\tau )$ of 1:1:1 \citep{PhysRevD.95.123001}, and search a maximum time window of 1 year. This upper limit is derived using the observed TS value and pseudo-experiments with added signal, assuming that the baseline Monte Carlo can be used to represent the signal. The median signal time-integrated flux required for a $5\sigma$ discovery, referred to as the discovery potential, is also shown in Figure~\ref{fig:ULs}. Comparing to AT2019dsg at $14.2^\circ$ declination, the upper limit is a factor of 2.5 above current model predictions \citep{winter_interpretation_2023} for neutrino emission from TDEs.

\section{stacking search}\label{sec:stacking}
The single source hypothesis outlined in Section \ref{sec:ss_methods} can be expanded to include an ensemble of sources, known as a stacking search. The stacking searches implemented in this work test the hypotheses that the astrophysical source classes of non-jetted and on-axis jetted TDEs emit neutrinos. By stacking small contributions from a number of TDEs, we can test for a collective signal above the background that may not be currently visible from individual sources.

\subsection{Analysis methodology}\label{sec:stackingmethod}
For a multi-source hypothesis, we use the same general form of the likelihood presented in Eq.~\ref{llh}, with changes added to the signal PDF to account for the source population and the relative contribution of each source. Here, we fit $n_s$ and $\gamma$ for each subpopulation, and fix $\Delta t$ to a 1 year time window with $t_0$ set to the UV or X-ray peak. To account for the source population in the likelihood, we sum the signal PDFs over each source in the stacking catalog. To consider the relative contribution of each source, we account for IceCube's variation in effective area across the sky, the time window for each source, and that sources do not contribute equally to the overall flux. We assume that all sources share the same intrinsic neutrino spectrum of a single power law and that each $k$-th source contributes a fraction $f_k$ to the overall flux from the complete ensemble of sources. Here, the standard candle assumption that each source's flux contribution is proportional to its luminosity distance, $D_L$ is used:
\begin{equation}
    f_k = \frac{1/D_{L,k}^2}{\sum_{k=1}^{N_{src}} (1 / D_{L,k}^2)}. 
\end{equation}
Using the $TS$ defined in Eq.~\ref{llh}, we calculate a $p$-value following the methods described in Section~\ref{sec:ss_methods}. For this stacking analysis, we test two separate hypotheses, one for non-jetted TDEs and one for on-axis jetted TDEs.  

\subsection{Stacking Search Results}
The results from each piece of the stacking test are consistent with the background-only hypothesis, as shown in Table~\ref{tab:stackingtable}. Upper limits are calculated at the 90\% confidence level for each subcatalog for an injected window of 150 days and an energy spectra of $E_\nu^{-2}$. The per-flavor upper limits reported in Table~\ref{tab:stackingtable} represent the average fluence per source in the catalog required to produce detectable stacking signal, weighted by the $1/D_L^2$ distance weighting scheme described in Section~\ref{sec:stackingmethod}. Rather than assuming a single source distance, the fluence limit reflects the effective contribution of a source at the flux-weighted mean distance of the catalog, accounting for the fact that nearby sources contribute more to the stacking signal than distant ones. 

\begin{figure*}[ht!]
\centering
\includegraphics[width=0.85\linewidth]{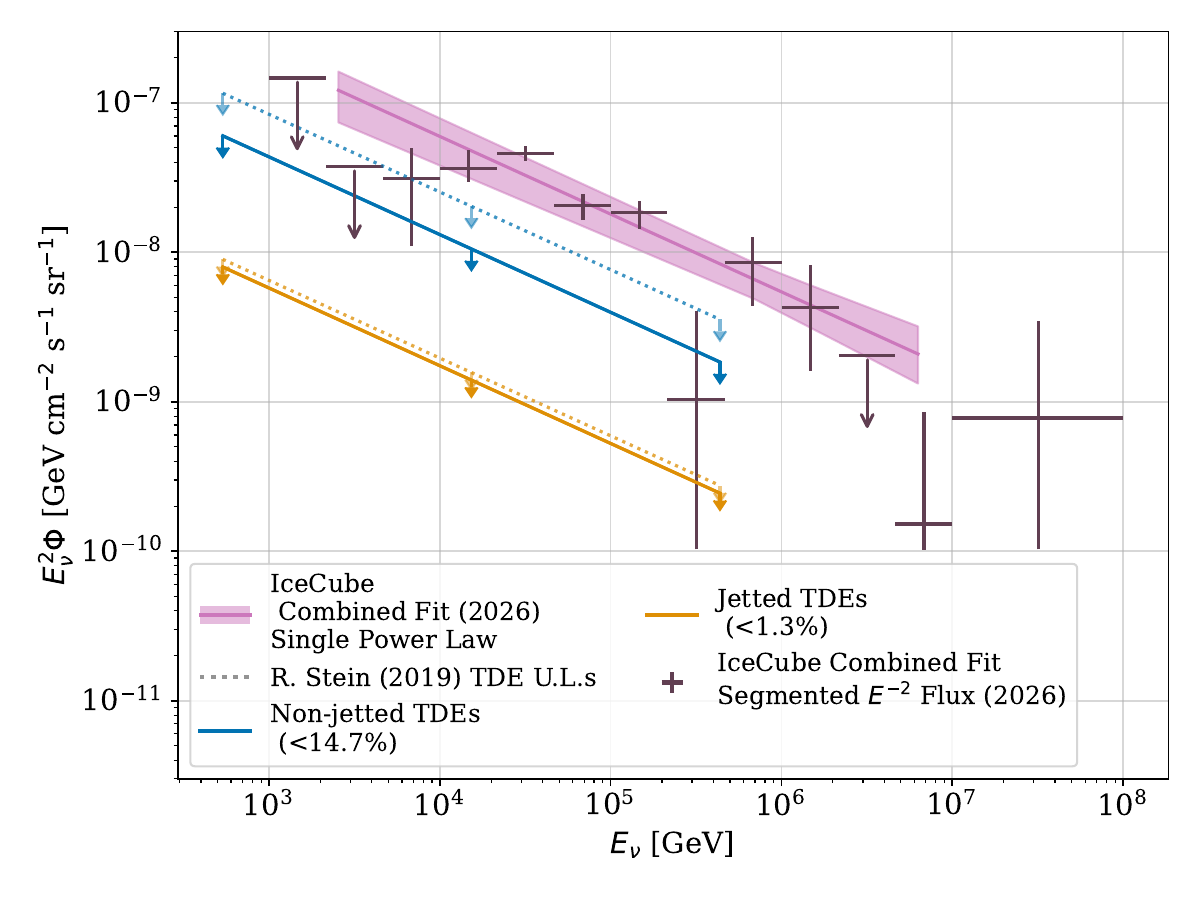}
\caption{\textbf{Diffuse flux contributions.} Limits on the per-flavor $\nu + \bar{\nu}$ contribution of jetted and non-jetted TDEs to the diffuse neutrino flux as a function of neutrino energy. The dashed lines show the comparison to previous analyses, recalculated to compare to the same diffuse flux, with that for jetted TDEs in orange and non-jetted TDEs in blue. 
\label{fig:diffuse}}
\end{figure*}

\begin{deluxetable*}{ccccccc}
\tablecolumns{7}
\tablewidth{0pt}
\tablecaption{Best fit $n_s$, $\gamma$, reported $p$-value, significance, and 90\% confidence level per-flavor $\nu+\bar{\nu}$ time-integrated flux upper limit results from the stacking searches for an injected window of 150 days and an energy spectra of $E_\nu^{-2}$, and assuming a flavor
ratio $(\nu_e:\nu_\mu:\nu_\tau )$ of 1:1:1 with equal contributions from $\nu$ and $\bar{\nu}$. The upper limit represents the average per-source flux weighted by the $1/D_L^2$ stacking weights. This limit can be rescaled to an equivalent limit for a source at distance $d$ by multiplying the flux by $(d_{eff}/d)^2$, where $d_{eff}$ is the effective average distance of the sources in the catalog. \label{tab:stackingtable}}
\tablehead{
\colhead{ }\vspace{-0.2cm} &  &  &  &  & \colhead{90\% C.L. U.L.} \\ 
\colhead{Subcategory}& \colhead{$n_s$} \vspace{-0.2cm} & \colhead{$\gamma$} & \colhead{$p$-value} & \colhead{Significance} & \colhead{\vspace{1.0cm}[GeV/cm$^2$]}
}
\startdata
Jetted & 2.5 & 2.5 & 0.31 & 0.49$\sigma$ & 5.3 $\times 10^{-2}$ \\
Non-jetted & 15.4 & 3.5 & 0.15 & 1.0$\sigma$ & 2.1 $ \times 10^{-1}$ \\
\enddata
\end{deluxetable*}
Unable to reject the null hypothesis, we set upper limits on contributions to the extragalactic diffuse neutrino flux. By using the assumption that TDEs behave as standard candles and that the corresponding flux on Earth is proportional to the inverse distance squared of each source, we can derive source class limits on neutrino emission. In reality, TDE populations are expected to span a range of intrinsic neutrino luminosities correlated with physical parameters such as the black hole mass and accretion rate. We inject an $E^{-2.52}$ astrophysical flux, from recent measurements of the astrophysical diffuse flux \citep{Abbasi_etal._2025}, uniformly across the search windows for each source and assume that the total emitted energy in neutrinos is the same for each source. We assume the TDE source evolution with redshift in \citep{sun_extragalactic_2015}. It is important to note that this evolution model was developed prior to the substantial increase in TDE observations enabled by wide field-of-view surveys, and therefore does not reflect the constraints that the current generation of TDE observations could place on the source density rate evolution. Limits have determined that this assumed rate underestimates the rate evolution with redshift \citep{Necker_2025}, but there is currently no proper measurement. The uncertainty in the source density rate evolution thus represents a significant uncertainty in the derived diffuse flux limits. Astrophysical transient populations are characterized by their local rate and their source evolution. Assuming the upper value of rate estimates of $8_{-4}^{+4} \times 10^{-7}$ Mpc$^{-3}$ yr$^{-1}$ for non-jetted TDEs \citep{s_van_velzen_mass_2018} and $3_{-2}^{+4} \times 10^{-11} $ Mpc$^-3$ yr$^{-1}$ for jetted TDEs \citep{hui_sun_extragalactic_2015}, we estimate the cumulative neutrino flux arising from each subpopulation, shown in Figure~\ref{fig:diffuse}. Figure~\ref{fig:diffuse} also compares the diffuse contributions calculated for this analysis with that of the 2019 analysis \citep{robert_stein_search_2019}, recalculated with the assumption of an $E^{-2.52}$ energy spectrum \citep{Abbasi_etal._2025}. We find that non-jetted and jetted TDEs contribute less than 14.7\% and 1.3\% to the diffuse neutrino flux, respectively, though these limits carry uncertainty from the assumed source evolution model and the standard candle approximation. 

\section{Discussion of Realtime Coincident Detections}\label{sec:realtime} 
The IceCube alert system started operating in 2016. In 2019, the alert system underwent an upgrade to improve track-like event selection \citep{abbasi_icecat-1_2023}, and was enhanced with a revised reconstruction strategy in 2024 to apply different algorithms depending on the event's deposited energy or topology \citep{icecube_icecat-2_2025}. Although \cite{robert_stein_tidal_2020} reported the TDE to be in coincidence with a high-energy neutrino reported by IceCube's realtime alert system, the recent improved reconstruction of the neutrino position found the TDE position to be well outside the updated neutrino containment region \citep{icecube_icecat-2_2025}. This analysis is consistent with this result, as tidal disruption event AT2019dsg was not found to be significant in this analysis. Furthermore, an analysis of high-energy neutrinos from TDE-like flares with IceCube reported no neutrinos observed from the coincident detections \citep{Abbasi:20239c}. As discussed in Section~\ref{sec:introduction}, the other optically detected accretion flares coincident with neutrino alerts \citep{reusch_multi-messenger_2023, vanvelzen_establishing_2024} were not included in this search due to uncertainties in their classification \citep{pitik_is_2022, guolo_agn_2023, sniegowska_at_2025}. The positions of these flares were also found to be outside the updated containment regions for the neutrino events \citep{icecube_icecat-2_2025}.

\section{Conclusions \& Future Outlook}\label{sec:conclusions}
This analysis focused on TDEs detected in optical/UV and X-rays. Both single source and stacking searches were conducted to look for evidence of neutrino emission from a dataset of 89 TDEs detected in optical/UV and X-rays. In the single source search, the most significant source after trials correction was AT2018fyk with a significance of $1.06\sigma$. Our upper limit for AT2019dsg is a factor of 2.5 above model predictions \citep{winter_interpretation_2023}. We set upper limits on the contributions to the diffuse high-energy neutrino flux for the stacking on-axis jetted and non-jetted TDE populations, which contribute less than 1.3\% and 14.7\% respectively, assuming the rates in \cite{hui_sun_extragalactic_2015} and \cite{s_van_velzen_mass_2018}. These upper limits are derived from the per-flavor $\nu + \bar{\nu}$ emission under the assumption of an unbroken power law evaluated over the energy range of the dataset used in this analysis, between $\sim 0.5$ TeV and $\sim 400$ TeV. If the neutrino luminosity function of TDEs is broad, the upper limits derived here could either overestimate or underestimate the true population constraint depending on whether nearby bright sources or distant faint sources dominate the stacking signal. Similarly, the assumption of a shared spectral index $\gamma = 2.5$ across all sources in a subpopulation may not hold if different TDE subclasses favor different neutrino production mechanisms with distinct spectral signatures. It is also important to note that the catalog used in this analysis is subject to the detection biases of the surveys from which it was drawn. If the neutrino-emitting TDE population is systematically under-represented in these samples, the stacking constraints presented here may not be representative of the full population. Analyzing TDEs and other accretion flares detected in infrared as potential sources of astrophysical neutrinos could be a promising future extension to this work.

The Legacy Survey of Space and Time (LSST) with the Vera C. Rubin Observatory \citep{ivezic_lsst_2019} is projected to detect $> 3000$ TDEs per year \citep{bricman_prospects_2020}, under the assumption of a detection requirement of 2 mag above the median $5\sigma$ limit. While the effective gain for neutrino source searches will depend on the redshift distribution of detected sources, the additional coverage in the southern sky combined with recent machine learning TDE classification efforts \citep{stein_tdescore_2024}, optimized multiwavelength data infrastructure \citep{otter}, and the improved potential to discover neutrino sources of IceCube-Gen2 \citep{aartsen_icecube-gen2_2021} could provide opportunities for improved population studies on TDEs as neutrino emitters.

\begin{acknowledgments}
The IceCube Collaboration acknowledges the significant contributions to this manuscript from Shannon Gray. The authors gratefully acknowledge the support from the following agencies and institutions: USA – U.S. National Science Foundation-Office of Polar Programs, U.S. National Science Foundation-Physics Division, U.S. National Science Foundation-EPSCoR, U.S. National Science Foundation-Office of Advanced Cyberinfrastructure, Wisconsin Alumni Research Foundation, Center for High Throughput Computing (CHTC) at the University of Wisconsin–Madison, Open Science Grid (OSG), Partnership to Advance Throughput Computing (PATh), Advanced Cyberinfrastructure Coordination Ecosystem: Services \& Support (ACCESS), Frontera and Ranch computing project at the Texas Advanced Computing Center, U.S. Department of Energy-National Energy Research Scientific Computing Center, Particle astrophysics research computing center at the University of Maryland, Michigan State University, Astroparticle physics computational facility at Marquette University, NVIDIA Corporation, and Google Cloud Platform; Belgium – Funds for Scientific Research (FRS-FNRS and FWO), FWO Odysseus and Big Science programmes, and Belgian Federal Science Policy Office (Belspo); Germany – Bundesministerium f\"ur Forschung, Technologie und Raumfahrt (BMFTR), Deutsche Forschungsgemeinschaft (DFG), Helmholtz Alliance for Astroparticle Physics (HAP), Initiative and Networking Fund of the Helmholtz Association, Deutsches Elektronen Synchrotron (DESY), and High Performance Computing cluster of the RWTH Aachen; Sweden – Swedish Research Council, Swedish Polar Research Secretariat,  National Academic Infrastructure for Supercomputing in Sweden (NAISS), and Knut and Alice Wallenberg Foundation; European Union – EGI Advanced Computing for research; Australia – Australian Research Council; Canada – Natural Sciences and Engineering Research Council of Canada, Calcul Qu\'{e}bec, Compute Ontario, Canada Foundation for Innovation, WestGrid, and Digital Research Alliance of Canada; Denmark – Villum Fonden, Carlsberg Foundation, and European Commission; New Zealand – Marsden Fund; Japan – Japan Society for Promotion of Science (JSPS), Ministry of Education, Culture, Sports, Science and Technology (MEXT), and Institute for Global Prominent Research (IGPR) of Chiba University; Korea – National Research Foundation of Korea (NRF); Switzerland – Swiss National Science Foundation (SNSF).
\end{acknowledgments}

\appendix

\section{Full catalog of sources}\label{a:sourcecat}
The source catalog was constructed by compiling spectroscopically confirmed TDEs detected in optical/UV or X-ray wavebands, drawing primarily from systematic ZTF TDE samples \citep{velzen_seventeen_2021, hammerstein_final_2022} supplemented by individual classifications from other surveys. Sources were included if they had a spectroscopically confirmed TDE classification, a well-defined electromagnetic peak date to serve as the reference time $T_0$, and a measured redshift. The catalog includes 34 TDEs with both X-ray and optical observations, 52 detected only optically, and 3 only in X-ray. 

TDEs selected from optical detections feature a light curve that follows a power-law decline from peak consistent with the theoretical stellar debris fallback rate \citep{gezari_tidal_2021}. TDEs selected in soft X-rays feature a soft spectrum well-fitted by blackbody temperatures in the range 0.02-0.13 keV with dramatic variability over time \citep{mummery_optical_2025}. While these selections use different observational criteria, both are representative of the overall unified TDE population \citep{guolo_systematic_2024}. Sources with significant classification ambiguity were excluded, including AT2019fdr and AT2019aalc, which have alternative classifications as a superluminous supernova \citep{pitik_is_2022} and AGN flare \citep{guolo_agn_2023}, respectively. The three jetted TDEs (Swift J1644+57, Swift J2058+05, Swift J112.2-82) are classified based on their confirmed relativistic jet signatures and analyzed separately from the non-jetted population in the stacking analysis. Generally, the exclusion of sources with strong pre-existing AGN activity is inherited from the classification criteria applied by the surveys and follow-up programs from which the catalog is drawn, which typically assess AGN contamination through prior photometric history, emission line diagnostics, and the spectral and temporal properties of the flare. The complete catalog with source information is shown in Table~\ref{tab:catalog}.

The catalog presented here is complementary to but distinct from the recently released data repository OTTER (Open mulTiwavelength Transient Event Repository) \citep{otter}, which provides a publicly available photometric archive of 240 TDE candidates with multiwavelength observations spanning radio to X-ray wavelengths. While OTTER is optimized for storing and accessing multiwavelength photometric datasets across the broadest possible sample of TDE candidates, the catalog used in this analysis required spectroscopic confirmation of the TDE classification, a well-defined electromagnetic peak date to anchor the neutrino emission time window, and for that peak to fall within May 2011 to November 2022, such that the time falls within available IceCube data stopping a year before the end of available data to uniformly search a year long time window. These selections resulted in a smaller but more uniformly characterized sample of 89 TDEs. 

\startlongtable
\begin{deluxetable*}{cccccccc}\label{tab:catalog}
\tablecolumns{8}
\tablecaption{TDE Catalog with jetted vs non-jetted categorization, right ascension and declination in degrees, peak date MJD, redshift $z$, and classification reference.}
\tablehead{ \colhead{} & \colhead{Type} & \colhead{Source Name} & \colhead{R.A. [$^\circ$]} & \colhead{Dec. [$^\circ$]} & \colhead{Peak Date} &  \colhead{$z$} &  \colhead{Ref.}}
\startdata
1 & PTF-09ge & Non-jetted & 14.95 & 49.61 & 54958 & 0.064 & \cite{arcavi_continuum_2014} \\
2 & PTF-09axc & Non-jetted & 223.30 & 22.24 & 55002 & 0.115 &  \cite{arcavi_continuum_2014} \\
3 & PTF-09djl & Non-jetted & 248.48 & 30.24 & 55036 & 0.184 &  \cite{arcavi_continuum_2014} \\
4 & PS1-10jh & Non-jetted & 242.37 & 53.67 & 55389 & 0.170 &  \cite{guillochon_ps1-10jh_2014}\\ 
5 & SDSS J1201+30 & Non-jetted & 180.40 & 30.05 & 55475 & 0.146 &   \cite{saxton_tidal_2012} \\ 
6 & PS1-11af & Non-jetted & 149.36 & 3.23 & 55581 & 0.405 & \cite{chornock_uv-bright_2013} \\
7 & SwiftJ1644+57 & Jetted & 251.21 & 57.58 & 55648 & 0.353 &  \cite{bloom_possible_2011}\\
8 & SwiftJ2058+05 & Jetted & 314.58 & 5.23 & 55698 & 1.186 &   \cite{bloom_possible_2011} \\
9 & SwiftJ1112.2-82 & Jetted & 167.95 & -82.65 & 55724 & 0.890 &  \cite{bloom_possible_2011} \\
10 & ASSASN-14ae & Non-jetted & 167.17 & 34.10 & 56682 & 0.044 &\cite{holoien_asassn-14ae_2014} \\
11 & ASSASN-14li & Non-jetted & 192.06 & 17.77 & 57015 & 0.021 &   \cite{holoien_six_2016} \\ 
12 & ASSASN-15lh & Non-jetted & 330.56 & -61.66 & 57178 & 0.233 & \cite{dong_asassn-15lh_2016} \\ 
13 & ASSASN-15oi & Non-jetted & 309.79 & -30.76 & 57248 & 0.048 &  \cite{holoien_six_2016} \\ 
14 & iPTF-15af & Non-jetted & 132.12 & 22.06 & 57370 & 0.079 &  \cite{blagorodnova_broad_2019} \\
15 & OGLE16aaa & Non-jetted & 16.84 & -64.27 & 57407 & 0.166 &  \cite{kajava_rapid_2020} \\ 
16 & iPTF-16axa & Non-jetted & 255.89 & 30.59 & 57482 & 0.108 &  \cite{hung_revisiting_2017} \\
17 & AT2016fnl/iPTF-16fnl & Non-jetted & 7.49 & 32.89 & 57630 & 0.016 &  \cite{blagorodnova_iptf16fnl_2017} \\
18 & AT2017eqx/PS17dhz & Non-jetted & 336.70 & 17.15 & 57921 & 0.109 &  \cite{nicholl_outflow_2020} \\
19 & AT2018meh/ATLAS18bcno & Non-jetted & 175.04 & 15.33 & 58136 & 0.011 &  \cite{johansson_ztf_2023} \\
20 & AT2018zr/ZTF18aabtxvd & Non-jetted & 119.23 & 34.26 & 58205 & 0.071  & \cite{van_velzen_first_2019}\\
21 & AT2018bsi/ZTF18aahqkbt & Non-jetted & 123.86 & 45.59 & 58217 & 0.051  & \cite{velzen_seventeen_2021} \\ 
22 & AT2018dyb/ASASSN-18pg & Non-jetted & 242.74 & -60.92 & 58340 & 0.018 &  \cite{leloudas_spectral_2019} \\
23 & AT2018fyk/ASASSN-18ul & Non-jetted & 342.57 & -44.86 & 58369 & 0.059  & \cite{wevers_evidence_2019} \\
24 & AT2018hco/ZTF18abxftqm & Non-jetted & 16.89 & 23.48 & 58403 & 0.088 & \cite{velzen_seventeen_2021}\\
25 & AT2018hyz/ZTF18acpdvos & Non-jetted & 151.71 & 1.69 & 58428 & 0.046 & \cite{gomez_tidal_2020}\\
26 & AT2018iih/ZTF18acaqdaa & Non-jetted & 262.02 & 30.69 & 58450 & 0.212 & \cite{velzen_seventeen_2021} \\
27 & AT2018lni/ZTF18actaqdw & Non-jetted & 62.41 & 73.89 & 58462 & 0.138 &  \cite{velzen_seventeen_2021}\\
28 & AT2018jbv/ZTF18acnbpmd & Non-jetted & 197.69 & 8.57 & 58469 & 0.340 &  \cite{hammerstein_final_2022} \\
29 & AT2018lna/ ZTF19aabbnzo & Non-jetted & 105.83 & 23.03 & 58508 & 0.091 & \cite{van_velzen_classification_2019}\\
30 & AT2019bhf/ZTF19aakswrb & Non-jetted & 227.32 & 16.24 & 58542 & 0.121 & \cite{dahiwale_ztf_2020}\\
31 & AT2019ahk/ASASSN-19bt & Non-jetted & 105.05 & -66.04 & 58546 & 0.026  & \cite{holoien_discovery_2019} \\
32 & AT2019cho/ZTF19aakiwze & Non-jetted & 193.79 & 49.52 & 58547 & 0.193 & \cite{velzen_seventeen_2021}\\
33 & AT2019azh/ZTF17aaazdba & Non-jetted & 123.32 & 22.65 & 58558 & 0.022  & \cite{velzen_seventeen_2021}\\
34 & AT2019dsg/ZTF19aapreis & Non-jetted & 314.26 & 14.20 & 58603 & 0.051  & \cite{velzen_seventeen_2021}\\ 
35 & AT2019eve/ZTF19aatylnl & Non-jetted & 172.21 & 15.67 & 58608 & 0.081 & \cite{mummery_optical_2025} \\ 
36 & AT2019ehz/ZTF19aarioci & Non-jetted & 212.42 & 55.49 & 58613 & 0.074  & \cite{velzen_seventeen_2021}\\
37 & AT2019gte/ZTF19aavxfib & Non-jetted & 196.66 & -1.55 & 58635 & 0.086 & \cite{swann_epessto_2019} \\ 
38 & AT2019meg/ZTF19abhhjcc & Non-jetted & 281.32 & 44.44 & 58688 & 0.152 &  \cite{velzen_seventeen_2021} \\
39 & AT2019lwu/ZTF19abidbya & Non-jetted & 347.80 & -1.00 & 58691 & 0.117 &  \cite{velzen_seventeen_2021}\\
40 & AT2019mha/ZTF19abhejal & Non-jetted & 244.12 & 56.43 & 58705 & 0.148 &  \cite{velzen_seventeen_2021}\\
41 & AT2019qiz/ZTF19abzrhgq & Non-jetted & 71.66 & -10.23 & 58763 & 0.015 &  \cite{velzen_seventeen_2021}\\
42 & AT2019teq/ZTF19accmaxo & Non-jetted & 284.77 & 47.52 & 58794 & 0.087  & \cite{hammerstein_final_2022} \\ 
43 & AT2019vcb/ZTF19acspeuw & Non-jetted & 189.73 & 33.17 & 58802 & 0.088  & \cite{dahiwale_ztf_2020} \\ 
44 & AT2020ddv/ZTF20aamqmfk & Non-jetted & 149.64 & 46.91 & 58812 & 0.160  & \cite{gezari_classification_2020}\\
45 & AT2020pj/ZTF20aabqihu & Non-jetted & 232.90 & 33.09 & 58829 & 0.068  & \cite{hammerstein_ztf_2020}\\
46 & SDSSJ1048+1228 & Non-jetted & 162.14 & 12.48 & 58977 & 0.054  & \cite{jian-he_zheng_choked_2023}\\
47 & AT2020ocn/ZTF18aakelin & Non-jetted & 208.47 & 54.00 & 58986 & 0.070 & \cite{cao_tidal_2024}\\ 
48 & AT2020neh/ZTF20abgwfek & Non-jetted & 230.33 & 14.07 & 59019 & 0.062 &  \cite{angus_fast-rising_2022}\\
49 & AT2020mbq/ZTF20abefeab & Non-jetted & 235.06 & 25.00 & 59020 & 0.093 &  \cite{hammerstein_final_2022}\\
50 & AT2020nov/ZTF20abisysx & Non-jetted & 254.55 & 2.12 & 59046 & 0.087  & \cite{hammerstein_final_2022} \\ 
51 & AT2020qhs/ZTF20abowque & Non-jetted & 34.47 & -9.61 & 59056 & 0.345 &  \cite{hammerstein_final_2022} \\
52 & SDSSJ1649+2625 & Non-jetted & 252.41 & 26.42 & 59068 & 0.059 & \cite{jian-he_zheng_choked_2023} \\
53 & AT2020ysg/ZTF20abnorit & Non-jetted & 171.36 & 27.44 & 59075 & 0.277 &  \cite{hammerstein_final_2022}\\
54 & AT2020riz/ZTF20abrnwfc & Non-jetted & 32.63 & 9.07 & 59080 & 0.435 &  \cite{hammerstein_final_2022}\\
55 & AT2020opy/ZTF20abjwvae & Non-jetted & 239.11 & 23.37 & 59089 & 0.159 &  \cite{goodwin_radio_2022}\\
56 & AT2020vdq/ZTF20acaazkt & Non-jetted & 152.22 & 42.72 & 59127 & 0.044  & \cite{mummery_optical_2025} \\
57 & AT2020vwl/ZTF20achpcvt & Non-jetted & 232.66 & 26.98 & 59132 & 0.033 &  \cite{hammerstein_ztf_2021-2} \\
58 & AT2020wey/ZTF20acitpfz & Non-jetted & 136.36 & 61.80 & 59150 & 0.027 & \cite{hammerstein_final_2022} \\
59 & AT2020yue/ZTF20acnznms & Non-jetted & 165.00 & 21.11 & 59156 & 0.204 & \cite{kangas_ztf_2022} \\
60 & AT2020zso/ZTF20acqoiyt & Non-jetted & 335.57 & -7.27 & 59156 & 0.057 &  \cite{ihanec_epessto_2020} \\
61 & AT2020acka/ZTF20acwytxn & Non-jetted & 238.76 & 16.30 & 59194 & 0.338 & \cite{hammerstein_final_2022} \\
62 & AT2021ack/ZTF20acyydkh & Non-jetted & 208.29 & 9.74 & 59203 & 0.133  & \cite{hammerstein_final_2022} \\ 
63 & AT2021lo/ZTF21aabiipy & Non-jetted & 46.11 & 4.75 & 59219 & 0.152 & \cite{yao_discovery_2022} \\ 
64 & AT2021jsg/ZTF21aaeoitd & Non-jetted & 167.15 & 30.76 & 59222 & 0.126 & \cite{yao_transient_2021} \\ 
65 & AT2021axu/ZTF21aaaokyp & Non-jetted & 176.65 & 30.09 & 59231 & 0.190 & \cite{hammerstein_ztf_2021-1} \\
66 & AT2021blz/ATLAS21djg & Non-jetted & 68.13 & -32.43 & 59246 & 0.045 & \cite{magee_epessto_2021}\\
67 & AT2021crk/ZTF21aakfqwq & Non-jetted & 176.28 & 18.54 & 59253 & 0.156 &  \cite{yuhan_yao_tidal_2023}\\
68 & AT2021ehb/ZTF21aanxhjv & Non-jetted & 46.95 & 40.31 & 59274 & 0.017 & \cite{gezari_ztf21aanxhjvat2021ehb_2021} \\
69 & AT2021gje/ZTF21aapvvtb & Non-jetted & 252.53 & 34.82 & 59292 & 0.358 & \cite{mummery_optical_2025} \\
70 & AT2021jjm/ZTF21aauuybx & Non-jetted & 219.88 & -27.86 & 59312 & 0.153 & \cite{yao_transient_2021}\\
71 & AT2021qxv/PS21grp & Non-jetted & 229.75 & -3.20 & 59344 & 0.183 & \cite{jones_ysepan-starrs1_2021} \\ 
72 & AT2021mhg/ZTF21abaxaqq & Non-jetted & 4.93 & 29.32 & 59345 & 0.073 & \cite{chu_ztf_2021}\\
73 & AT2021nwa/ZTF21abcgnqn & Non-jetted & 238.46 & 55.59 & 59345 & 0.047 &\cite{yao_ztf_2021}\\
74 & AT2021qth/ZTF21abhrchb & Non-jetted & 302.91 & -21.16 & 59383 & 0.081 & \cite{hammerstein_final_2022}\\
75 & AT2022gri/ZTF20aahmtso & Non-jetted & 109.59 & 33.99 & 59640 & 0.028 & \cite{yao_ztf20aahmtsoat2022gri_2022} \\
76 & AT2022lri/ZTF22abajudi & Non-jetted & 35.03 & -22.72 & 59732 & 0.032 & \cite{yuhan_yao_tidal_2023}\\
77 & AT2022rz/ZTF22aaaedas & Non-jetted & 234.49 & 56.27 & 59592 & 0.107 & \cite{hammerstein_ztf_2022} \\ 
78 & AT2022aee/ZTF22aaabovl & Non-jetted & 132.18 & 80.88 & 59600 & 0.124 & \cite{yao_discovery_2022} \\
79 & AT2022adm/ZTF22aaaekqh & Non-jetted & 216.96 & 28.17 & 59600 & 0.061 & \cite{munoz-arancibia_alerceztf_2022} \\ 
80 & AT2022arb/ZTF22aaaihgr & Non-jetted & 156.04 & -0.82 & 59607 & 0.062 & \cite{tonry_atlas_2022}\\
81 & AT2022bdw/ZTF22aaahtqz & Non-jetted & 126.29 & 18.58 & 59610 & 0.039 & \cite{anumarlapudi_radio_2024} \\ 
82 & AT2022dbl/ZTF18aabdajx & Non-jetted & 185.19 & 49.55 & 59632 & 0.028 & \cite{hinkle_double_2024} \\ 
83 & AT2022dyt/ZTF22aacgcwv & Non-jetted & 150.03 & 26.46 & 59636 & 0.072 & \cite{somalwar_transient_2022}\\ 
84 & AT2022exr/ZTF22aadgefj & Non-jetted & 262.46 & 25.84 & 59658 & 0.096 & \cite{guolo_delayed_2022}\\
85 & AT2022ibq/ZTF22aagvrlq & Non-jetted & 267.65 & 21.27 & 59670 & 0.395 & \cite{yao_discovery_2022}\\
86 & AT2022hvp/ZTF22aagyuao & Non-jetted & 148.69 & 55.44 & 59688 & 0.12 & \cite{fulton_classification_2022} \\ 
87 & AT2022pna/ZTF22aavvqyh & Non-jetted & 25.48 & -3.29 & 59784 & 0.095 &\cite{yuhan_yao_tidal_2023}\\
88 & AT2022upj/ZTF22abegjtx & Non-jetted & 5.99 & -14.42 & 59822 & 0.054 &\cite{chakraborty_discovery_2025} \\
89 & AT2022wtn/ZTF22abkfhua & Non-jetted & 350.85 & 10.69 & 59854 & 0.049 & \cite{fremling_ztf_2022}\\
\enddata
\end{deluxetable*}

\bibliography{bib}{}
\bibliographystyle{aasjournalv7}



\end{document}

%% file: authors.tex
\affiliation{III. Physikalisches Institut, RWTH Aachen University, D-52056 Aachen, Germany}
\affiliation{Department of Physics, University of Adelaide, Adelaide, 5005, Australia}
\affiliation{Dept. of Physics and Astronomy, University of Alaska Anchorage, 3211 Providence Dr., Anchorage, AK 99508, USA}
\affiliation{School of Physics and Center for Relativistic Astrophysics, Georgia Institute of Technology, Atlanta, GA 30332, USA}
\affiliation{Dept. of Physics, Southern University, Baton Rouge, LA 70813, USA}
\affiliation{Dept. of Physics, University of California, Berkeley, CA 94720, USA}
\affiliation{Lawrence Berkeley National Laboratory, Berkeley, CA 94720, USA}
\affiliation{Institut f{\"u}r Physik, Humboldt-Universit{\"a}t zu Berlin, D-12489 Berlin, Germany}
\affiliation{Fakult{\"a}t f{\"u}r Physik {\&} Astronomie, Ruhr-Universit{\"a}t Bochum, D-44780 Bochum, Germany}
\affiliation{Universit{\'e} Libre de Bruxelles, Science Faculty CP230, B-1050 Brussels, Belgium}
\affiliation{Vrije Universiteit Brussel (VUB), Dienst ELEM, B-1050 Brussels, Belgium}
\affiliation{Dept. of Physics, Simon Fraser University, Burnaby, BC V5A 1S6, Canada}
\affiliation{Department of Physics and Laboratory for Particle Physics and Cosmology, Harvard University, Cambridge, MA 02138, USA}
\affiliation{Dept. of Physics, Massachusetts Institute of Technology, Cambridge, MA 02139, USA}
\affiliation{Dept. of Physics and The International Center for Hadron Astrophysics, Chiba University, Chiba 263-8522, Japan}
\affiliation{Department of Physics, Loyola University Chicago, Chicago, IL 60660, USA}
\affiliation{Dept. of Physics and Astronomy, University of Canterbury, Private Bag 4800, Christchurch, New Zealand}
\affiliation{Dept. of Physics, University of Maryland, College Park, MD 20742, USA}
\affiliation{Dept. of Astronomy, Ohio State University, Columbus, OH 43210, USA}
\affiliation{Dept. of Physics and Center for Cosmology and Astro-Particle Physics, Ohio State University, Columbus, OH 43210, USA}
\affiliation{Niels Bohr Institute, University of Copenhagen, DK-2100 Copenhagen, Denmark}
\affiliation{Dept. of Physics, TU Dortmund University, D-44221 Dortmund, Germany}
\affiliation{Dept. of Physics and Astronomy, Michigan State University, East Lansing, MI 48824, USA}
\affiliation{Dept. of Physics, University of Alberta, Edmonton, Alberta, T6G 2E1, Canada}
\affiliation{Erlangen Centre for Astroparticle Physics, Friedrich-Alexander-Universit{\"a}t Erlangen-N{\"u}rnberg, D-91058 Erlangen, Germany}
\affiliation{Physik-department, Technische Universit{\"a}t M{\"u}nchen, D-85748 Garching, Germany}
\affiliation{D{\'e}partement de physique nucl{\'e}aire et corpusculaire, Universit{\'e} de Gen{\`e}ve, CH-1211 Gen{\`e}ve, Switzerland}
\affiliation{Dept. of Physics and Astronomy, University of Gent, B-9000 Gent, Belgium}
\affiliation{Dept. of Physics and Astronomy, University of California, Irvine, CA 92697, USA}
\affiliation{Karlsruhe Institute of Technology, Institute for Astroparticle Physics, D-76021 Karlsruhe, Germany}
\affiliation{Karlsruhe Institute of Technology, Institute of Experimental Particle Physics, D-76021 Karlsruhe, Germany}
\affiliation{Dept. of Physics, Engineering Physics, and Astronomy, Queen's University, Kingston, ON K7L 3N6, Canada}
\affiliation{Department of Physics {\&} Astronomy, University of Nevada, Las Vegas, NV 89154, USA}
\affiliation{Nevada Center for Astrophysics, University of Nevada, Las Vegas, NV 89154, USA}
\affiliation{Dept. of Physics and Astronomy, University of Kansas, Lawrence, KS 66045, USA}
\affiliation{UCLouvain, Centre for Cosmology, Particle Physics and Phenomenology, CP3, Chemin du Cyclotron 2, 1348 Louvain-la-Neuve, Belgium}
\affiliation{Department of Physics, Mercer University, Macon, GA 31207-0001, USA}
\affiliation{Dept. of Astronomy, University of Wisconsin{\textemdash}Madison, Madison, WI 53706, USA}
\affiliation{Dept. of Physics and Wisconsin IceCube Particle Astrophysics Center, University of Wisconsin{\textemdash}Madison, Madison, WI 53706, USA}
\affiliation{Institute of Physics, University of Mainz, Staudinger Weg 7, D-55099 Mainz, Germany}
\affiliation{Department of Physics, Marquette University, Milwaukee, WI 53201, USA}
\affiliation{Institut f{\"u}r Kernphysik, Universit{\"a}t M{\"u}nster, D-48149 M{\"u}nster, Germany}
\affiliation{Bartol Research Institute and Dept. of Physics and Astronomy, University of Delaware, Newark, DE 19716, USA}
\affiliation{Dept. of Physics, Yale University, New Haven, CT 06520, USA}
\affiliation{Columbia Astrophysics and Nevis Laboratories, Columbia University, New York, NY 10027, USA}
\affiliation{Dept. of Physics, University of Oxford, Parks Road, Oxford OX1 3PU, United Kingdom}
\affiliation{Dipartimento di Fisica e Astronomia Galileo Galilei, Universit{\`a} Degli Studi di Padova, I-35122 Padova PD, Italy}
\affiliation{Dept. of Physics, Drexel University, 3141 Chestnut Street, Philadelphia, PA 19104, USA}
\affiliation{Physics Department, South Dakota School of Mines and Technology, Rapid City, SD 57701, USA}
\affiliation{Dept. of Physics, University of Wisconsin, River Falls, WI 54022, USA}
\affiliation{Dept. of Physics and Astronomy, University of Rochester, Rochester, NY 14627, USA}
\affiliation{Department of Physics and Astronomy, University of Utah, Salt Lake City, UT 84112, USA}
\affiliation{Dept. of Physics, Chung-Ang University, Seoul 06974, Republic of Korea}
\affiliation{Oskar Klein Centre and Dept. of Physics, Stockholm University, SE-10691 Stockholm, Sweden}
\affiliation{Dept. of Physics and Astronomy, Stony Brook University, Stony Brook, NY 11794-3800, USA}
\affiliation{Dept. of Physics, Sungkyunkwan University, Suwon 16419, Republic of Korea}
\affiliation{Institute of Physics, Academia Sinica, Taipei, 11529, Taiwan}
\affiliation{Dept. of Physics and Astronomy, University of Alabama, Tuscaloosa, AL 35487, USA}
\affiliation{Dept. of Astronomy and Astrophysics, Pennsylvania State University, University Park, PA 16802, USA}
\affiliation{Dept. of Physics, Pennsylvania State University, University Park, PA 16802, USA}
\affiliation{Dept. of Physics and Astronomy, Uppsala University, Box 516, SE-75120 Uppsala, Sweden}
\affiliation{Dept. of Physics, University of Wuppertal, D-42119 Wuppertal, Germany}
\affiliation{Deutsches Elektronen-Synchrotron DESY, Platanenallee 6, D-15738 Zeuthen, Germany}

\author[0000-0001-6141-4205]{R. Abbasi}
\affiliation{Department of Physics, Loyola University Chicago, Chicago, IL 60660, USA}
\email{rabbasi@luc.edu}

\author[0000-0001-8952-588X]{M. Ackermann}
\affiliation{Deutsches Elektronen-Synchrotron DESY, Platanenallee 6, D-15738 Zeuthen, Germany}
\email{markus.ackermann@desy.de}

\author{J. Adams}
\affiliation{Dept. of Physics and Astronomy, University of Canterbury, Private Bag 4800, Christchurch, New Zealand}
\email{jenni.adams@canterbury.ac.nz}

\author[0000-0003-2252-9514]{J. A. Aguilar}
\affiliation{Universit{\'e} Libre de Bruxelles, Science Faculty CP230, B-1050 Brussels, Belgium}
\email{juanan.aguilar@icecube.wisc.edu}

\author[0000-0003-0709-5631]{M. Ahlers}
\affiliation{Niels Bohr Institute, University of Copenhagen, DK-2100 Copenhagen, Denmark}
\email{mahlers@icecube.wisc.edu}

\author[0000-0002-9534-9189]{J.M. Alameddine}
\affiliation{Dept. of Physics, TU Dortmund University, D-44221 Dortmund, Germany}
\email{jean-marco.alameddine@icecube.wisc.edu}

\author[0009-0001-2444-4162]{S. Ali}
\affiliation{Dept. of Physics and Astronomy, University of Kansas, Lawrence, KS 66045, USA}
\email{shoukat@ku.edu}

\author{N. M. Amin}
\affiliation{Bartol Research Institute and Dept. of Physics and Astronomy, University of Delaware, Newark, DE 19716, USA}
\email{moureen@udel.edu}

\author[0000-0001-9394-0007]{K. Andeen}
\affiliation{Department of Physics, Marquette University, Milwaukee, WI 53201, USA}
\email{karen.andeen@icecube.wisc.edu}

\author[0000-0003-4186-4182]{C. Arg{\"u}elles}
\affiliation{Department of Physics and Laboratory for Particle Physics and Cosmology, Harvard University, Cambridge, MA 02138, USA}
\email{carlos.arguelles@icecube.wisc.edu}

% ...

\date{\today}

\collaboration{420}{IceCube Collaboration}

%% file: bib.bib
@article{ji_at2022sxl_2025,
	title = {{AT2022sxl}: {A} {Candidate} {Repeating} {Tidal} {Disruption} {Event} in {Possible} {Association} with {Two} {High}-energy {Neutrino} {Events}},
	volume = {991},
	issn = {0004-637X},
	shorttitle = {{AT2022sxl}},
	url = {https://doi.org/10.3847/1538-4357/adfb73},
	doi = {10.3847/1538-4357/adfb73},
	language = {en},
	number = {1},
	urldate = {2025-10-31},
	journal = {The Astrophysical Journal},
	author = {Ji, Shunhao and Wang, Zhongxiang and Zhu, Litao and Geier, Stefan and Gupta, Alok C.},
	month = sep,
	year = {2025},
	note = {Publisher: The American Astronomical Society},
	pages = {20},
}

@article{yuan__at2021lwx_2024,
	title = {{AT2021lwx}: {Another} {Neutrino}-coincident {Tidal} {Disruption} {Event} with a {Strong} {Dust} {Echo}?},
	volume = {969},
	issn = {0004-637X, 1538-4357},
	shorttitle = {{AT2021lwx}},
	url = {https://iopscience.iop.org/article/10.3847/1538-4357/ad50a9},
	doi = {10.3847/1538-4357/ad50a9},
	number = {2},
	urldate = {2025-10-31},
	journal = {The Astrophysical Journal},
	author = {Yuan, Chengchao and Winter, Walter and Lunardini, Cecilia},
	month = jul,
	year = {2024},
	pages = {136},
}

@article{Dai_Fang_2017, title={Can tidal disruption events produce the IceCube neutrinos?}, volume={469}, ISSN={0035-8711, 1365-2966}, DOI={10.1093/mnras/stx863}, number={2}, journal={Monthly Notices of the Royal Astronomical Society}, author={Dai, Lixin and Fang, Ke}, year={2017}, month=aug, pages={1354–1359}, language={en} }

@article{gezari_tidal_2021,
	title = {Tidal {Disruption} {Events}},
	volume = {59},
	issn = {1545-4282},
	url = {https://www.annualreviews.org/content/journals/10.1146/annurev-astro-111720-030029},
	doi = {https://doi.org/10.1146/annurev-astro-111720-030029},
	number = {Volume 59, 2021},
	journal = {Annual Review of Astronomy and Astrophysics},
	author = {Gezari, Suvi},
	year = {2021},
	note = {Publisher: Annual Reviews
Type: Journal Article},
	pages = {21--58},
}

@article{guolo_systematic_2024,
	title = {A {Systematic} {Analysis} of the {X}-{Ray} {Emission} in {Optically} {Selected} {Tidal} {Disruption} {Events}: {Observational} {Evidence} for the {Unification} of the {Optically} and {X}-{Ray}-selected {Populations}},
	volume = {966},
	url = {https://iopscience.iop.org/article/10.3847/1538-4357/ad2f9f},
	doi = {10.3847/1538-4357/ad2f9f},
	number = {160},
	journal = {The Astrophysical Journal},
	author = {Guolo, Muryel and {Suvi Gezari} and {Yuhan Yao} and {Sjoert van Velzen}},
	year = {2024},
}

@article{hui_sun_extragalactic_2015,
	title = {Extragalactic {High}-{Energy} {Transients}: {Event} {Rate} {Densities} and {Luminosity} {Functions}},
	volume = {812},
	url = {https://iopscience.iop.org/article/10.1088/0004-637X/812/1/33},
	doi = {10.1088/0004-637X/812/1/33},
	number = {33},
	journal = {The Astrophysical Journal},
	author = {Hui Sun and Bing Zhang and Zhuo Li},
	year = {2015},
}

@article{jian-he_zheng_choked_2023,
	title = {Choked {Jets} in {Expanding} {Envelope} as the {Origin} of the {Neutrino} {Emission} {Associated} with {Tidal} {Disruption} {Events}},
	volume = {954},
	url = {https://iopscience.iop.org/article/10.3847/1538-4357/ace71c},
	number = {17},
	journal = {The Astrophysical Journal},
	author = {{Jian-He Zheng} and {Ruo-Yu Liu} and {Xiang-Yu Wang}},
	month = aug,
	year = {2023},
}

@article{robert_stein_search_2019,
	title = {Search for {Neutrinos} from {Populations} of {Optical} {Transients}},
	doi = {https://doi.org/10.22323/1.358.1016},
	journal = {Proceedings of Science},
	author = {R. Stein and Aartsen, M. G. and Abbasi, R. and Abdou, Y. and Ackermann, M. and Adams, J. and Aguilar, J. A. and Ahlers, M. and Altmann, D. and Auffenberg, J. and Bai, X. and Baker, M. and Barwick, S. W. and Baum, V. and Bay, R. and Beatty, J. J. and Bechet, S. and Tjus, J. Becker and Becker, K.-H. and Benabderrahmane, M. L. and BenZvi, S. and Berghaus, P.},
	month = aug,
	year = {2019},
	note = {ICRC 2019},
}

@article{Wevers_2023, title={Live to Die Another Day: The Rebrightening of AT 2018fyk as a Repeating Partial Tidal Disruption Event}, volume={942}, ISSN={2041-8205, 2041-8213}, DOI={10.3847/2041-8213/ac9f36}, abstractNote={Abstract
            
              Stars that interact with supermassive black holes (SMBHs) can be either completely or partially destroyed by tides. In a partial tidal disruption event (TDE), the high-density core of the star remains intact, and the low-density outer envelope of the star is stripped and feeds a luminous accretion episode. The TDE AT 2018fyk, with an inferred black hole mass of 10
              7.7±0.4
              M
              ⊙
              , experienced an extreme dimming event at X-ray (factor of >6000) and UV (factor of ∼15) wavelengths ∼500–600 days after discovery. Here we report on the reemergence of these emission components roughly 1200 days after discovery. We find that the source properties are similar to those of the predimming accretion state, suggesting that the accretion flow was rejuvenated to a similar state. We propose that a repeated partial TDE, where the partially disrupted star is on an ∼1200 day orbit about the SMBH and periodically stripped of mass during each pericenter passage, powers its unique light curve. This scenario provides a plausible explanation for AT 2018fyk’s overall properties, including the rapid dimming event and the rebrightening at late times. We also provide testable predictions for the behavior of the accretion flow in the future; if the second encounter was also a partial disruption, then we predict another strong dimming event around day 1800 (2023 August) and a subsequent rebrightening around day 2400 (2025 March). This source provides strong evidence of the partial disruption of a star by an SMBH.}, number={2}, journal={The Astrophysical Journal Letters}, author={Wevers, T. and Coughlin, E. R. and Pasham, D. R. and Guolo, M. and Sun, Y. and Wen, S. and Jonker, P. G. and Zabludoff, A. and Malyali, A. and Arcodia, R. and Liu, Z. and Merloni, A. and Rau, A. and Grotova, I. and Short, P. and Cao, Z.}, year={2023}, month=jan, pages={L33} }

@misc{otter,
      title={The Open mulTiwavelength Transient Event Repository (OTTER): Infrastructure Release and Tidal Disruption Event Catalog}, 
      author={Noah Franz and Kate D Alexander and Sebastian Gomez and Collin T Christy and Tanmoy Laskar and Sjoert van Velzen and Nicholas Earl and Suvi Gezari and Mitchell Karmen and Raffaella Margutti and Jeniveve Pearson and V. Ashley Villar and Ann I Zabludoff},
      year={2026},
      eprint={2509.05405},
      archivePrefix={arXiv},
      primaryClass={astro-ph.HE},
      url={https://arxiv.org/abs/2509.05405}, 
}

@article{s_van_velzen_mass_2018,
	title = {On the {Mass} and {Luminosity} {Functions} of {Tidal} {Disruption} {Flares}: {Rate} {Suppression} due to {Black} {Hole} {Event} {Horizons}},
	volume = {852},
	doi = {10.3847/1538-4357/aa998e},
	number = {72},
	journal = {The Astrophysical Journal},
	author = {S. van Velzen},
	month = jan,
	year = {2018},
	note = {This article is corrected by 2018 ApJ 868 154 (https://iopscience.iop.org/article/10.3847/1538-4357/aae9da)},
}

@article{robert_stein_tidal_2020,
	title = {A tidal disruption event coincident with a high-energy neutrino},
	volume = {5},
	doi = {https://doi.org/10.1038/s41550-020-01295-8},
	journal = {Nature Astronomy},
	author = {R. Stein and Sjoert van Velzen and Marek Kowalski and Anna Frackowiak and {Suvi Gezari} and {James C. A. Miller-Jones} and {Sara Frederick} and {Itai Sfaradi}},
	month = jul,
	year = {2020},
	pages = {510--518},
}

@article{yuhan_yao_tidal_2023,
	title = {Tidal {Disruption} {Event} {Demographics} with the {Zwicky} {Transient} {Facility}: {Volumetric} {Rates}, {Luminosity} {Function}, and {Implications} for the {Local} {Black} {Hole} {Mass} {Function}},
	volume = {955},
	doi = {10.3847/2041-8213/acf216},
	number = {L6},
	journal = {The Astrophysical Journal Letters},
	author = {{Yuhan Yao} and {Vikram Ravi} and {Suvi Gezari} and {Sjoert van Velzen} and {Wenbin Lu} and {Steve Schulze} and {Jean J. Somalwar} and {Erica Hammerstein}},
	month = sep,
	year = {2023},
}

@article{winter_interpretation_2023,
	title = {Interpretation of the observed neutrino emission from three {Tidal} {Disruption} {Events}},
	volume = {948},
	issn = {0004-637X, 1538-4357},
	url = {http://arxiv.org/abs/2205.11538},
	doi = {10.3847/1538-4357/acbe9e},
	number = {1},
	urldate = {2025-08-11},
	journal = {The Astrophysical Journal},
	author = {Winter, Walter and Lunardini, Cecilia},
	month = may,
	year = {2023},
	note = {arXiv:2205.11538 [astro-ph]},
	pages = {42},
}

@article{velzen_seventeen_2021,
	title = {Seventeen {Tidal} {Disruption} {Events} from the {First} {Half} of {ZTF} {Survey} {Observations}: {Entering} a {New} {Era} of {Population} {Studies}},
	volume = {908},
	issn = {0004-637X, 1538-4357},
	shorttitle = {Seventeen {Tidal} {Disruption} {Events} from the {First} {Half} of {ZTF} {Survey} {Observations}},
	url = {http://arxiv.org/abs/2001.01409},
	doi = {10.3847/1538-4357/abc258},
	number = {1},
	urldate = {2025-08-11},
	journal = {The Astrophysical Journal},
	author = {van Velzen, Sjoert  and Gezari, Suvi and Hammerstein, Erica and Roth, Nathaniel and Frederick, Sara and Ward, Charlotte and Hung, Tiara and Cenko, S. Bradley and Stein, Robert and Perley, Daniel A. and Taggart, Kirsty and Sollerman, Jesper and Andreoni, Igor and Bellm, Eric C. and Brinnel, Valery and De, Kishalay and Dekany, Richard and Feeney, Michael and Foley, Ryan J. and Fremling, Christoffer and Giomi, Matteo and Golkhou, V. Zach and Ho, Anna Y. Q. and Kasliwal, Mansi M. and Kilpatrick, Charles D. and Kulkarni, Shrinivas R. and Kupfer, Thomas and Laher, Russ R. and Mahabal, Ashish and Masci, Frank J. and Nordin, Jakob and Riddle, Reed and Rusholme, Ben and Sharma, Yashvi and Santen, Jakob van and Shupe, David L. and Soumagnac, Maayane T.},
	month = feb,
	year = {2021},
	note = {arXiv:2001.01409 [astro-ph]},
	pages = {4},
}

@article{senno_high-energy_2017,
	title = {High-{Energy} {Neutrino} {Flares} {From} {X}-{Ray} {Bright} and {Dark} {Tidal} {Disruptions} {Events}},
	volume = {838},
	issn = {1538-4357},
	url = {http://arxiv.org/abs/1612.00918},
	doi = {10.3847/1538-4357/aa6344},
	number = {1},
	urldate = {2025-08-11},
	journal = {The Astrophysical Journal},
	author = {Senno, Nicholas and Murase, Kohta and Meszaros, Peter},
	month = mar,
	year = {2017},
	note = {arXiv:1612.00918 [astro-ph]},
	pages = {3},
}

@misc{reusch_multi-messenger_2023,
	title = {Multi-messenger {Observations} of {Tidal} {Disruption} {Events}},
	url = {http://arxiv.org/abs/2307.00902},
	doi = {10.48550/arXiv.2307.00902},
	urldate = {2025-08-11},
	publisher = {arXiv},
	author = {Reusch, Simeon},
	month = jul,
	year = {2023},
	note = {arXiv:2307.00902 [astro-ph]},
}

@article{murase_high-energy_2019,
	title = {High-{Energy} {Multimessenger} {Transient} {Astrophysics}},
	volume = {69},
	issn = {0163-8998, 1545-4134},
	url = {https://www.annualreviews.org/content/journals/10.1146/annurev-nucl-101918-023510},
	doi = {10.1146/annurev-nucl-101918-023510},
	language = {en},
	number = {Volume 69, 2019},
	urldate = {2025-09-18},
	journal = {Annual Review of Nuclear and Particle Science},
	author = {Murase, Kohta and Bartos, Imre},
	month = oct,
	year = {2019},
	note = {Publisher: Annual Reviews},
	pages = {477--506},
}

@article{farrar_giant_2009,
	title = {Giant {AGN} {Flares} and {Cosmic} {Ray} {Bursts}},
	volume = {693},
	doi = {10.1088/0004-637X/693/1/329},
	journal = {Astrophys. J.},
	author = {Farrar, Glennys R. and Gruzinov, Andrei},
	year = {2009},
	note = {\_eprint: 0802.1074},
	pages = {329--332},
}

@article{wang_extreme_2012,
	title = {{EXTREME} {CORONAL} {LINE} {EMITTERS}: {TIDAL} {DISRUPTION} {OF} {STARS} {BY} {MASSIVE} {BLACK} {HOLES} {IN} {GALACTIC} {NUCLEI}?},
	volume = {749},
	issn = {0004-637X},
	shorttitle = {{EXTREME} {CORONAL} {LINE} {EMITTERS}},
	url = {https://dx.doi.org/10.1088/0004-637X/749/2/115},
	doi = {10.1088/0004-637X/749/2/115},
	language = {en},
	number = {2},
	urldate = {2025-09-18},
	journal = {The Astrophysical Journal},
	author = {Wang, Ting-Gui and Zhou, Hong-Yan and Komossa, S. and Wang, Hui-Yuan and Yuan, Weimin and Yang, Chenwei},
	month = mar,
	year = {2012},
	note = {Publisher: The American Astronomical Society},
	pages = {115},
}

@article{guepin_ultra-high-energy_2018,
	title = {Ultra-high-energy cosmic rays and neutrinos from tidal disruptions by massive black holes},
	volume = {616},
	copyright = {© ESO 2018},
	issn = {0004-6361, 1432-0746},
	url = {https://www.aanda.org/articles/aa/abs/2018/08/aa32392-17/aa32392-17.html},
	doi = {10.1051/0004-6361/201732392},
	language = {en},
	urldate = {2025-09-18},
	journal = {Astronomy \& Astrophysics},
	author = {Guépin, Claire and Kotera, Kumiko and Barausse, Enrico and Fang, Ke and Murase, Kohta},
	month = aug,
	year = {2018},
	note = {Publisher: EDP Sciences},
	pages = {A179},
}

@article{pitik_is_2022,
	title = {Is the {High}-energy {Neutrino} {Event} {IceCube}-{200530A} {Associated} with a {Hydrogen}-rich {Superluminous} {Supernova}?},
	volume = {929},
	issn = {0004-637X},
	url = {https://dx.doi.org/10.3847/1538-4357/ac5ab1},
	doi = {10.3847/1538-4357/ac5ab1},
	language = {en},
	number = {2},
	urldate = {2025-09-18},
	journal = {The Astrophysical Journal},
	author = {Pitik, Tetyana and Tamborra, Irene and Angus, Charlotte R. and Auchettl, Katie},
	month = apr,
	year = {2022},
	note = {Publisher: The American Astronomical Society},
	pages = {163},
}

@article{vanvelzen_establishing_2024,
	title = {Establishing accretion flares from supermassive black holes as a source of high-energy neutrinos},
	volume = {529},
	issn = {0035-8711},
	url = {https://doi.org/10.1093/mnras/stae610},
	doi = {10.1093/mnras/stae610},
	number = {3},
	urldate = {2025-09-18},
	journal = {Monthly Notices of the Royal Astronomical Society},
	author = {van Velzen, Sjoert and Stein, Robert and Gilfanov, Marat and Kowalski, Marek and Hayasaki, Kimitake and Reusch, Simeon and Yao, Yuhan and Garrappa, Simone and Franckowiak, Anna and Gezari, Suvi and Nordin, Jakob and Fremling, Christoffer and Sharma, Yashvi and Yan, Lin and Kool, Erik C and Stern, Daniel and Veres, Patrik M and Sollerman, Jesper and Medvedev, Pavel and Sunyaev, Rashid and Bellm, Eric C and Dekany, Richard G and Duev, Dimitri A and Graham, Matthew J and Kasliwal, Mansi M and Kulkarni, Shrinivas R and Laher, Russ R and Riddle, Reed L and Rusholme, Ben},
	month = apr,
	year = {2024},
	pages = {2559--2576},
}

@article{sniegowska_at_2025,
	title = {{AT} 2019aalc: {A} {Bowen} {Fluorescence} {Flare} with a {Precursor} {Flare} in an {Active} {Galactic} {Nucleus}},
	volume = {989},
	issn = {0004-637X},
	shorttitle = {{AT} 2019aalc},
	url = {https://dx.doi.org/10.3847/1538-4357/aded13},
	doi = {10.3847/1538-4357/aded13},
	language = {en},
	number = {2},
	urldate = {2025-09-18},
	journal = {The Astrophysical Journal},
	author = {\'Sniegowska, Marzena and Trakhtenbrot, Benny and Makrygianni, Lydia and Arcavi, Iair and Ricci, Claudio and Faris, Sara and Palit, Biswaraj and Howell, D. Andrew and Newsome, Megan and Farah, Joseph and McCully, Curtis and Padilla-Gonzalez, Estefania and Terreran, Giacomo},
	month = aug,
	year = {2025},
	note = {Publisher: The American Astronomical Society},
	pages = {173},
}

@article{guolo_agn_2023,
	title = {On the {AGN} nature of {AT2019aalc}/{ZTF19aaejtoy}},
	volume = {195},
	url = {https://ui.adsabs.harvard.edu/abs/2023TNSAN.195....1G},
	urldate = {2025-09-18},
	journal = {Transient Name Server AstroNote},
	author = {Guolo, M. and Gezari, S.},
	month = jul,
	year = {2023},
	note = {ADS Bibcode: 2023TNSAN.195....1G},
	pages = {1},
}

@article{sun_extragalactic_2015,
	title = {{EXTRAGALACTIC} {HIGH}-{ENERGY} {TRANSIENTS}: {EVENT} {RATE} {DENSITIES} {AND} {LUMINOSITY} {FUNCTIONS}},
	volume = {812},
	issn = {0004-637X},
	shorttitle = {{EXTRAGALACTIC} {HIGH}-{ENERGY} {TRANSIENTS}},
	url = {https://dx.doi.org/10.1088/0004-637X/812/1/33},
	doi = {10.1088/0004-637X/812/1/33},
	language = {en},
	number = {1},
	urldate = {2025-09-18},
	journal = {The Astrophysical Journal},
	author = {Sun, Hui and Zhang, Bing and Li, Zhuo},
	month = oct,
	year = {2015},
	note = {Publisher: The American Astronomical Society},
	pages = {33},
}

@article{piran_disk_2015,
	title = {‧{Disk} {Formation} {Versus} {Disk} {Accretion}—{What} {Powers} {Tidal} {Disruption} {Events}?},
	volume = {806},
	issn = {0004-637X},
	url = {https://ui.adsabs.harvard.edu/abs/2015ApJ...806..164P},
	doi = {10.1088/0004-637X/806/2/164},
	urldate = {2025-09-18},
	journal = {The Astrophysical Journal},
	author = {Piran, Tsvi and Svirski, Gilad and Krolik, Julian and Cheng, Roseanne M. and Shiokawa, Hotaka},
	month = jun,
	year = {2015},
	note = {ADS Bibcode: 2015ApJ...806..164P},
	pages = {164},
}

@article{roth_radiative_2020,
	title = {Radiative {Emission} {Mechanisms}},
	volume = {216},
	issn = {0038-6308, 1572-9672},
	url = {https://link.springer.com/10.1007/s11214-020-00735-1},
	doi = {10.1007/s11214-020-00735-1},
	language = {en},
	number = {7},
	urldate = {2025-09-18},
	journal = {Space Science Reviews},
	author = {Roth, Nathaniel and Rossi, Elena Maria and Krolik, Julian and Piran, Tsvi and Mockler, Brenna and Kasen, Daniel},
	month = oct,
	year = {2020},
	pages = {114},
}

@article{hammerstein_final_2022,
	title = {The {Final} {Season} {Reimagined}: 30 {Tidal} {Disruption} {Events} from the {ZTF}-{I} {Survey}},
	volume = {942},
	issn = {0004-637X},
	shorttitle = {The {Final} {Season} {Reimagined}},
	url = {https://dx.doi.org/10.3847/1538-4357/aca283},
	doi = {10.3847/1538-4357/aca283},
	language = {en},
	number = {1},
	urldate = {2025-09-24},
	journal = {The Astrophysical Journal},
	author = {Hammerstein, Erica and van Velzen, Sjoert and Gezari, Suvi and Cenko, S. Bradley and Yao, Yuhan and Ward, Charlotte and Frederick, Sara and Villanueva, Natalia and Somalwar, Jean J. and Graham, Matthew J. and Kulkarni, Shrinivas R. and Stern, Daniel and Andreoni, Igor and Bellm, Eric C. and Dekany, Richard and Dhawan, Suhail and Drake, Andrew J. and Fremling, Christoffer and Gatkine, Pradip and Groom, Steven L. and Ho, Anna Y. Q. and Kasliwal, Mansi M. and Karambelkar, Viraj and Kool, Erik C. and Masci, Frank J. and Medford, Michael S. and Perley, Daniel A. and Purdum, Josiah and Roestel, Jan van and Sharma, Yashvi and Sollerman, Jesper and Taggart, Kirsty and Yan, Lin},
	month = dec,
	year = {2022},
	note = {Publisher: The American Astronomical Society},
	pages = {9},
}

@article{ivezic_lsst_2019,
	title = {{LSST}: {From} {Science} {Drivers} to {Reference} {Design} and {Anticipated} {Data} {Products}},
	volume = {873},
	issn = {0004-637X, 1538-4357},
	shorttitle = {{LSST}},
	url = {https://iopscience.iop.org/article/10.3847/1538-4357/ab042c},
	doi = {10.3847/1538-4357/ab042c},
	number = {2},
	urldate = {2025-10-17},
	journal = {The Astrophysical Journal},
	author = {Ivezić, {\v Z}eljko and Kahn, Steven M. and Tyson, J. Anthony and Abel, Bob and Acosta, Emily and Allsman, Robyn and Alonso, David and AlSayyad, Yusra and Anderson, Scott F. and Andrew, John and P. Angel, James Roger and Angeli, George Z. and Ansari, Reza and Antilogus, Pierre and Araujo, Constanza and Armstrong, Robert and Arndt, Kirk T. and Astier, Pierre and Aubourg, Éric and Auza, Nicole and Axelrod, Tim S. and Bard, Deborah J. and Barr, Jeff D. and Barrau, Aurelian and Bartlett, James G. and Bauer, Amanda E. and Bauman, Brian J. and Baumont, Sylvain and Bechtol, Ellen and Bechtol, Keith and Becker, Andrew C. and Becla, Jacek and Beldica, Cristina and Bellavia, Steve and Bianco, Federica B. and Biswas, Rahul and Blanc, Guillaume and Blazek, Jonathan and Blandford, Roger D. and Bloom, Josh S. and Bogart, Joanne and Bond, Tim W. and Booth, Michael T. and Borgland, Anders W. and Borne, Kirk and Bosch, James F. and Boutigny, Dominique and Brackett, Craig A. and Bradshaw, Andrew and Brandt, William Nielsen and Brown, Michael E. and Bullock, James S. and Burchat, Patricia and Burke, David L. and Cagnoli, Gianpietro and Calabrese, Daniel and Callahan, Shawn and Callen, Alice L. and Carlin, Jeffrey L. and Carlson, Erin L. and Chandrasekharan, Srinivasan and Charles-Emerson, Glenaver and Chesley, Steve and Cheu, Elliott C. and Chiang, Hsin-Fang and Chiang, James and Chirino, Carol and Chow, Derek and Ciardi, David R. and Claver, Charles F. and Cohen-Tanugi, Johann and Cockrum, Joseph J. and Coles, Rebecca and Connolly, Andrew J. and Cook, Kem H. and Cooray, Asantha and Covey, Kevin R. and Cribbs, Chris and Cui, Wei and Cutri, Roc and Daly, Philip N. and Daniel, Scott F. and Daruich, Felipe and Daubard, Guillaume and Daues, Greg and Dawson, William and Delgado, Francisco and Dellapenna, Alfred and Peyster, Robert De and Val-Borro, Miguel De and Digel, Seth W. and Doherty, Peter and Dubois, Richard and Dubois-Felsmann, Gregory P. and Durech, Josef and Economou, Frossie and Eifler, Tim and Eracleous, Michael and Emmons, Benjamin L. and Neto, Angelo Fausti and Ferguson, Henry and Figueroa, Enrique and Fisher-Levine, Merlin and Focke, Warren and Foss, Michael D. and Frank, James and Freemon, Michael D. and Gangler, Emmanuel and Gawiser, Eric and Geary, John C. and Gee, Perry and Geha, Marla and Gessner, Charles J. B. and Gibson, Robert R. and Gilmore, D. Kirk and Glanzman, Thomas and Glick, William and Goldina, Tatiana and Goldstein, Daniel A. and Goodenow, Iain and Graham, Melissa L. and Gressler, William J. and Gris, Philippe and Guy, Leanne P. and Guyonnet, Augustin and Haller, Gunther and Harris, Ron and Hascall, Patrick A. and Haupt, Justine and Hernandez, Fabio and Herrmann, Sven and Hileman, Edward and Hoblitt, Joshua and Hodgson, John A. and Hogan, Craig and Howard, James D. and Huang, Dajun and Huffer, Michael E. and Ingraham, Patrick and Innes, Walter R. and Jacoby, Suzanne H. and Jain, Bhuvnesh and Jammes, Fabrice and Jee, M. James and Jenness, Tim and Jernigan, Garrett and Jevremović, Darko and Johns, Kenneth and Johnson, Anthony S. and Johnson, Margaret W. G. and Jones, R. Lynne and Juramy-Gilles, Claire and Jurić, Mario and Kalirai, Jason S. and Kallivayalil, Nitya J. and Kalmbach, Bryce and Kantor, Jeffrey P. and Karst, Pierre and Kasliwal, Mansi M. and Kelly, Heather and Kessler, Richard and Kinnison, Veronica and Kirkby, David and Knox, Lloyd and Kotov, Ivan V. and Krabbendam, Victor L. and Krughoff, K. Simon and Kubánek, Petr and Kuczewski, John and Kulkarni, Shri and Ku, John and Kurita, Nadine R. and Lage, Craig S. and Lambert, Ron and Lange, Travis and Langton, J. Brian and Guillou, Laurent Le and Levine, Deborah and Liang, Ming and Lim, Kian-Tat and Lintott, Chris J. and Long, Kevin E. and Lopez, Margaux and Lotz, Paul J. and Lupton, Robert H. and Lust, Nate B. and MacArthur, Lauren A. and Mahabal, Ashish and Mandelbaum, Rachel and Markiewicz, Thomas W. and Marsh, Darren S. and Marshall, Philip J. and Marshall, Stuart and May, Morgan and McKercher, Robert and McQueen, Michelle and Meyers, Joshua and Migliore, Myriam and Miller, Michelle and Mills, David J. and Miraval, Connor and Moeyens, Joachim and Moolekamp, Fred E. and Monet, David G. and Moniez, Marc and Monkewitz, Serge and Montgomery, Christopher and Morrison, Christopher B. and Mueller, Fritz and Muller, Gary P. and Arancibia, Freddy Muñoz and Neill, Douglas R. and Newbry, Scott P. and Nief, Jean-Yves and Nomerotski, Andrei and Nordby, Martin and O’Connor, Paul and Oliver, John and Olivier, Scot S. and Olsen, Knut and O’Mullane, William and Ortiz, Sandra and Osier, Shawn and Owen, Russell E. and Pain, Reynald and Palecek, Paul E. and Parejko, John K. and Parsons, James B. and Pease, Nathan M. and Peterson, J. Matt and Peterson, John R. and Petravick, Donald L. and Petrick, M. E. Libby and Petry, Cathy E. and Pierfederici, Francesco and Pietrowicz, Stephen and Pike, Rob and Pinto, Philip A. and Plante, Raymond and Plate, Stephen and Plutchak, Joel P. and Price, Paul A. and Prouza, Michael and Radeka, Veljko and Rajagopal, Jayadev and Rasmussen, Andrew P. and Regnault, Nicolas and Reil, Kevin A. and Reiss, David J. and Reuter, Michael A. and Ridgway, Stephen T. and Riot, Vincent J. and Ritz, Steve and Robinson, Sean and Roby, William and Roodman, Aaron and Rosing, Wayne and Roucelle, Cecille and Rumore, Matthew R. and Russo, Stefano and Saha, Abhijit and Sassolas, Benoit and Schalk, Terry L. and Schellart, Pim and Schindler, Rafe H. and Schmidt, Samuel and Schneider, Donald P. and Schneider, Michael D. and Schoening, William and Schumacher, German and Schwamb, Megan E. and Sebag, Jacques and Selvy, Brian and Sembroski, Glenn H. and Seppala, Lynn G. and Serio, Andrew and Serrano, Eduardo and Shaw, Richard A. and Shipsey, Ian and Sick, Jonathan and Silvestri, Nicole and Slater, Colin T. and Smith, J. Allyn and Smith, R. Chris and Sobhani, Shahram and Soldahl, Christine and Storrie-Lombardi, Lisa and Stover, Edward and Strauss, Michael A. and Street, Rachel A. and Stubbs, Christopher W. and Sullivan, Ian S. and Sweeney, Donald and Swinbank, John D. and Szalay, Alexander and Takacs, Peter and Tether, Stephen A. and Thaler, Jon J. and Thayer, John Gregg and Thomas, Sandrine and Thornton, Adam J. and Thukral, Vaikunth and Tice, Jeffrey and Trilling, David E. and Turri, Max and Berg, Richard Van and Berk, Daniel Vanden and Vetter, Kurt and Virieux, Francoise and Vucina, Tomislav and Wahl, William and Walkowicz, Lucianne and Walsh, Brian and Walter, Christopher W. and Wang, Daniel L. and Wang, Shin-Yawn and Warner, Michael and Wiecha, Oliver and Willman, Beth and Winters, Scott E. and Wittman, David and Wolff, Sidney C. and Wood-Vasey, W. Michael and Wu, Xiuqin and Xin, Bo and Yoachim, Peter and Zhan, Hu},
	month = mar,
	year = {2019},
	pages = {111},
}

@article{Necker_2025, title={Flaires: A comprehensive catalog of dust echo-like infrared flares}, volume={695}, rights={https://creativecommons.org/licenses/by/4.0}, ISSN={0004-6361, 1432-0746}, DOI={10.1051/0004-6361/202451340}, abstractNote={Context.
              Observations of transient emission from extreme accretion events onto supermassive black holes can reveal conditions in the center of galaxies and the black hole itself. Most recently, it has been suggested these sources could be emitters of high-energy neutrinos. However, in most cases, it remains unclear whether this would be classified as the outcome of rejuvenated accretion or a tidal disruption event (TDE).
            
            
              Aims.
              We expand on existing samples of infrared (IR) flares to compile the largest and most complete list available. A large sample size is necessary to provide high-enough statistics for distant and faint objects to estimate their rates. Our catalog is large enough to facilitate a preliminary study of the rate evolution with redshift for the first time.
            
            
              Methods.
              We compiled a sample of 40 million galaxies. Using a custom, publicly available pipeline, we analyzed the WISE light curves for these 40 million objects using the Bayesian Blocks algorithm. We selected promising for dust echo candidates involved in transient accretion events and we inferred the luminosity, extension, and temperature of the hot dust by fitting a blackbody spectrum.
            
            
              Results.
              We established a clean sample of 823 dust echo-like IR flares, dubbed the Flaires catalog. For 568 of them, we were able to estimate the dust properties. After removing 70 objects with possible contributions from synchrotron emission, the luminosity, extension, and temperature are consistent with dust echos. Estimating the dust extension from the light curve shape revealed that the duration of the incident flare is broadly compatible with the duration of TDEs. The resulting rate per galaxy is consistent with the latest measurements of IR-detected TDEs and appears to decline with increasing redshift.
            
            
              Conclusions.
              Although systematic uncertainties may impact the calculation of the rate evolution, this catalog will enable further research of phenomena related to dust echos from TDEs and extreme accretion flares.}, journal={Astronomy \& Astrophysics}, author={Necker, J. and Graikou, E. and Kowalski, M. and Franckowiak, A. and Nordin, J. and Pernice, T. and Van Velzen, S. and Veres, P. M.}, year={2025}, month=mar, pages={A228} }

@article{Abbasi:20239c,
  author = "Abbasi, Rasha  and  Ackermann, Markus  and  Adams, Jenni  and  Agarwalla, Sanjib Kumar  and  Aguilar, Juanan  and  Ahlers, Markus  and  Alameddine, Jean-Marco  and  Amin, Najia Moureen Binte  and  Andeen, Karen  and  Anton, Gisela  and  Argüelles, Carlos  and  Ashida, Yosuke  and  Athanasiadou, Sofia  and  Axani, Spencer  and  Bai, Xinhua  and  Balagopal V., Aswathi  and  Baricevic, Moreno  and  Barwick, Steve  and  Basu, Vedant  and  Bay, Ryan  and  Beatty, James  and  Becker Tjus, Julia  and  Beise, Jakob  and  Bellenghi, Chiara  and  Benning, Charlotte  and  BenZvi, Segev  and  Berley, David  and  Bernardini, Elisa  and  Besson, Dave  and  Blaufuss, Erik  and  Blot, Summer  and  Bontempo, Federico  and  Book, Julia  and  Boscolo Meneguolo, Caterina  and  BOSER, Sebastian  and  Botner, Olga  and  Bottcher, Jakob  and  Bourbeau, Etienne  and  Braun, Jim  and  Brinson, Bennett  and  Brostean-Kaiser, Jannes  and  Burley, Ryan T.  and  Busse, Raffaela  and  Butterfield, Delaney  and  Campana, Michael  and  Carloni, Kiara  and  Carnie-Bronca, Erin  and  Chattopadhyay, Sharmistha  and  Chau, Thien Nhan  and  Chen, Chujie  and  Chen, Zheyang  and  Chirkin, Dmitry  and  Choi, Seowon  and  Clark, Brian  and  Classen, Lew  and  Coleman, Alan  and  Collin, Gabriel  and  Connolly, Amy  and  Conrad, Janet  and  Coppin, Paul  and  Correa, Pablo  and  Cowen, Doug  and  Dave, Pranav  and  DE CLERCQ, Catherine  and  DeLaunay, James  and  Delgado Lopez, Diyaselis  and  Deng, Shuyang  and  Deoskar, Kunal  and  Desai, Abhishek  and  Desiati, Paolo  and  de Vries, Krijn  and  de Wasseige, Gwenhaël  and  DeYoung, Tyce  and  Diaz, Alejandro  and  Diaz-Velez, Juan Carlos  and  Dittmer, Markus  and  Domi, Alba  and  Dujmovic, Hrvoje  and  DuVernois, Michael  and  Ehrhardt, Thomas  and  Eller, Philipp  and  Ellinger, Enrico  and  El Mentawi, Sharif  and  Elsässer, Dominik  and  Engel, Ralph  and  Erpenbeck, Hannah  and  Evans, John  and  Evenson, Paul  and  Fan, Kwok Lung  and  Fang, Ke  and  Farrag, Kareem Ramadan  and  Fazely, Ali  and  Fedynitch, Anatoli  and  Feigl, Nora  and  Fiedlschuster, Sebastian  and  Finley, Chad  and  Fischer, Leander  and  Fox, Derek B  and  Franckowiak, Anna  and  Fritz, Alexander  and  Furst, Philipp  and  Gallagher, Jay  and  Ganster, Erik  and  Garcia, Alfonso  and  Gerhardt, Lisa  and  Ghadimi, Ava  and  Glaser, Christian  and  Glauch, Theo  and  Glusenkamp, Thorsten  and  Goehlke, Noah  and  Gonzalez, Javier  and  Goswami, Sreetama  and  Grant, Darren  and  Gray, Shannon  and  Gries, Oliver  and  Griffin, Sean  and  Griswold, Spencer  and  Groth, Kathrine Morch  and  Günther, Christoph  and  Gutjahr, Pascal  and  Haack, Christian  and  Hallgren, Allan  and  Halliday, Robert  and  Halve, Lasse  and  Halzen, Francis  and  Hamdaoui, Hassane  and  Ha Minh, Martin  and  Hanson, Kael  and  Hardin, John  and  Harnisch, Alexander  and  Hatch, Patrick  and  Haungs, Andreas  and  Helbing, Klaus  and  Hellrung, Jonas  and  Henningsen, Felix  and  Heuermann, Lars Philipp  and  Heyer, Nils  and  Hickford, Stephanie  and  Hidvegi, Attila  and  Hill, Colton  and  Hill, Gary  and  Hoffman, Kara  and  Hori, Sam  and  Hoshina, Kotoyo  and  Hou, Wenjie  and  Huber, Thomas  and  Hultqvist, Klas  and  Hunnefeld, Mirco  and  Hussain, Raamis  and  Hymon, Karolin  and  In, Seongjin  and  Ishihara, Aya  and  Jacquart, Marc  and  Janik, Oliver  and  Jansson, Matti  and  Japaridze, George  and  Jeong, Minjin  and  Jin, Miaochen  and  Jones, Ben  and  Kang, Donghwa  and  Kang, Woosik  and  Kang, Xinyue  and  Kappes, Alexander  and  Kappesser, David  and  Kardum, Leonora  and  Karg, Timo  and  Karl, Martina  and  Karle, Albrecht  and  Katz, Ulì  and  Kauer, Matt  and  Kelley, John  and  Khatee Zathul, Arifa  and  Kheirandish, Ali  and  Kiryluk, Joanna  and  Klein, Spencer  and  Kochocki, Alina  and  Koirala, Ramesh  and  Kolanoski, Hermann  and  Kontrimas, Tomas  and  Kopke, Lutz  and  Kopper, Claudio  and  Koskinen, Jason  and  Koundal, Paras  and  Kovacevich, Michael  and  Kowalski, Marek  and  Kozynets, Tetiana  and  Jayakumar, Krishnamoorthi  and  Kruiswijk, Karlijn  and  Krupczak, Emmett  and  Kumar, Anil  and  Kun, Emma  and  Neilson, Naoko Kurahashi  and  Lad, Neha Navnitkumar  and  Lagunas Gualda, Cristina  and  Lamoureux, Mathieu  and  Larson, Michael J  and  Latseva, Silvia  and  Lauber, Frederik Hermann  and  Lazar, Jeffrey  and  Lee, Jiwoong  and  Leonard DeHolton, Kayla  and  Leszczynska, Agnieszka  and  Lincetto, Massimiliano  and  Liu, Qinrui  and  Liubarska, Maria  and  Lohfink, Elisa  and  Love, Christina  and  Lozano Mariscal, Cristian Jesus  and  Lu, Lu  and  Lucarelli, Francesco  and  Luszczak, William  and  Lyu, Yang  and  Madsen, Jim  and  Mahn, Kendall  and  Makino, Yuya  and  Manao, Elena  and  Mancina, Sarah  and  Marie Sainte, Wenceslas  and  Maris, Ioana Codrina  and  Marka, Szabolcs  and  Marka, Zsuzsa  and  Marsee, Matthew  and  Martinez-Soler, Ivan  and  Maruyama, Reina H.  and  Mayhew, Finn  and  McElroy, Thomas  and  McNally, Frank  and  Mead, James Vincent  and  Meagher, Kevin  and  Mechbal, Sarah  and  Medina, Andres  and  Meier, Maximilian  and  Merckx, Yarno  and  Merten, Lukas  and  Micallef, Jessie  and  Mitchell, Justin  and  Montaruli, Teresa  and  Moore, Roger  and  Morii, Yasutsugu  and  Morse, Bob  and  Moulai, Marjon  and  Mukherjee, Tista  and  Naab, Richard  and  Nagai, Ryo  and  Nakos, Maxwell  and  Naumann, Uwe  and  Necker, Jannis  and  Negi, Akshima  and  Neumann, Miriam  and  Niederhausen, Hans  and  Nisa, Mehr  and  Noell, Andreas  and  Novikov, Alexander  and  Nowicki, Sarah  and  Pollmann, Anna  and  O'Dell, Vivian  and  Oehler, Marie  and  Oeyen, Bob  and  Olivas, Alex  and  Ørsøe, Rasmus  and  Osborn, Jesse  and  O'Sullivan, Erin  and  Pandya, Hershal  and  Park, Nahee  and  Parker, Grant  and  Paudel, Ek Narayan  and  Paul, Larissa  and  Pérez de los Heros, Carlos  and  Peterson, Josh  and  Philippen, Saskia  and  Pizzuto, Alex  and  Plum, Matthias  and  Ponten, Axel  and  Popovych, Yuriy  and  Prado Rodriguez, Maria  and  Pries, Brandon  and  Procter-Murphy, Rachel  and  Przybylski, Gerald  and  Raab, Christoph  and  Rack-Helleis, John  and  Rawlins, Katherine  and  Rechav, Zoe  and  Rehman, Abdul  and  Reichherzer, Patrick  and  Renzi, Giovanni  and  Resconi, Elisa  and  Reusch, Simeon  and  Rhode, Wolfgang  and  Riedel, Benedikt  and  Rifaie, Adam  and  Roberts, Ella  and  Robertson, Sally  and  Rodan, Steven T.  and  Roellinghoff, Gerrit  and  Rongen, Martin  and  Rott, Carsten  and  Ruhe, Tim  and  Ruohan, Li  and  Ryckbosch, Dirk  and  Safa, Ibrahim  and  Saffer, Julian  and  Salazar-Gallegos, Daniel  and  Sampathkumar, Pranav  and  Sanchez Herrera, Sebastian  and  Sandrock, Alexander  and  Santander, Marcos  and  Sarkar, Sourav  and  Sarkar, Subir  and  Savelberg, Joelle  and  Savina, Pierpaolo  and  Schaufel, Merlin  and  Schieler, Harald  and  Schindler, Sebastian  and  Schlickmann, Lea  and  Schlüter, Berit  and  Schlüter, Felix  and  Schmeisser, Nick  and  Schmidt, Torsten  and  Schneider, Judith  and  Schröder, Frank  and  Schumacher, Lisa Johanna  and  Schwefer, Georg  and  Sclafani, Steve  and  Seckel, David  and  Seikh, Mohammad Ful Hossain  and  Seunarine, Suruj  and  Shah, Riya  and  Sharma, Ankur  and  Shefali, {}  and  Shimizu, Nobuhiro  and  Silva, Manuel  and  Skrzypek, Barbara  and  Smithers, Ben  and  Snihur, Robert  and  Soedingrekso, Jan  and  Sogaard, Andreas  and  Soldin, Dennis  and  Soldin, Philipp  and  Sommani, Giacomo  and  Spannfellner, Christian  and  Spiczak, Glenn  and  Spiering, Christian  and  Stamatikos, Michael  and  Stanev, Todor  and  Stezelberger, Thorsten  and  Sturwald, Timo  and  Stuttard, Thomas  and  Sullivan, Greg  and  Taboada, Ignacio  and  Ter-Antonyan, Samvel  and  Thiesmeyer, Matthias  and  Thompson, Will  and  Thwaites, Jessie  and  Tilav, Serap  and  Tollefson, Kirsten  and  Tönnis, Christoph  and  Toscano, Simona  and  Tosi, Delia  and  Trettin, Alexander  and  Tung, Chun Fai  and  Turcotte, Roxanne  and  Twagirayezu, Jean Pierre  and  Ty, Bunheng  and  Unland Elorrieta, Martin  and  Upadhyay, Anuj Kumar  and  Upshaw, Karriem  and  Valtonen-Mattila, Nora  and  Vandenbroucke, Justin  and  van Eijndhoven, Nick  and  Vannerom, David  and  van Santen, Jakob  and  Vara, Javi  and  Veitch-Michaelis, Joshua  and  Venugopal, Megha  and  Vereecken, Matthias  and  Verpoest, Stef  and  Veske, Doğa  and  Vijai, Aishwarya  and  Walck, Christian  and  Weaver, Chris  and  Weigel, Philip  and  Weindl, Andreas  and  Weldert, Jan  and  Wendt, Chris  and  Werthebach, Johannes  and  Weyrauch, Mark  and  Whitehorn, Nathan  and  Wiebusch, Christopher  and  Willey, Nathan  and  Williams, Dawn  and  Witthaus, Lucas  and  Wolf, Annika  and  Wolf, Martin  and  Wrede, Gerrit  and  Xu, Xianwu  and  Yanez, Juan Pablo  and  Yildizci, Emre Burak  and  Yoshida, Shigeru  and  Young, Robert  and  Yu, Felix J.  and  Yu, Shiqi  and  Yuan, Tianlu  and  Zhang, Zelong  and  Zhelnin, Pavel  and  Zimmerman, Melany",
  title = "{Search for High-Energy Neutrinos from TDE-like Flares with IceCube}",
  doi = "10.22323/1.444.1478",
  journal = "PoS",
  year = 2023,
  volume = "ICRC2023",
  pages = "1478"
}

@article{bricman_prospects_2020,
	title = {The {Prospects} of {Observing} {Tidal} {Disruption} {Events} with the {Large} {Synoptic} {Survey} {Telescope}},
	volume = {890},
	issn = {0004-637X, 1538-4357},
	url = {https://iopscience.iop.org/article/10.3847/1538-4357/ab6989},
	doi = {10.3847/1538-4357/ab6989},
	number = {1},
	urldate = {2025-10-20},
	journal = {The Astrophysical Journal},
	author = {Bricman, Katja and Gomboc, Andreja},
	month = feb,
	year = {2020},
	pages = {73},
}

@article{stein_tdescore_2024,
	title = {tdescore: {An} {Accurate} {Photometric} {Classifier} for {Tidal} {Disruption} {Events}},
	volume = {965},
	issn = {2041-8205, 2041-8213},
	shorttitle = {tdescore},
	url = {https://iopscience.iop.org/article/10.3847/2041-8213/ad3337},
	doi = {10.3847/2041-8213/ad3337},
	number = {2},
	urldate = {2025-10-20},
	journal = {The Astrophysical Journal Letters},
	author = {Stein, Robert and Mahabal, Ashish and Reusch, Simeon and Graham, Matthew and Kasliwal, Mansi M. and Kowalski, Marek and Gezari, Suvi and Hammerstein, Erica and Nakoneczny, Szymon J. and Nicholl, Matt and Sollerman, Jesper and van Velzen, Sjoert and Yao, Yuhan and Laher, Russ R. and Rusholme, Ben},
	month = apr,
	year = {2024},
	pages = {L14},
}

@article{blagorodnova_iptf16fnl_2017,
	title = {{iPTF16fnl}: {A} {Faint} and {Fast} {Tidal} {Disruption} {Event} in an {E}+{A} {Galaxy}},
	volume = {844},
	issn = {0004-637X, 1538-4357},
	shorttitle = {{iPTF16fnl}},
	url = {https://iopscience.iop.org/article/10.3847/1538-4357/aa7579},
	doi = {10.3847/1538-4357/aa7579},
	number = {1},
	urldate = {2025-10-21},
	journal = {The Astrophysical Journal},
	author = {Blagorodnova, N. and Gezari, S. and Hung, T. and Kulkarni, S. R. and Cenko, S. B. and Pasham, D. R. and Yan, L. and Arcavi, I. and Ben-Ami, S. and Bue, B. D. and Cantwell, T. and Cao, Y. and Castro-Tirado, A. J. and Fender, R. and Fremling, C. and Gal-Yam, A. and Ho, A. Y. Q. and Horesh, A. and Hosseinzadeh, G. and Kasliwal, M. M. and Kong, A. K. H. and Laher, R. R. and Leloudas, G. and Lunnan, R. and Masci, F. J. and Mooley, K. and Neill, J. D. and Nugent, P. and Powell, M. and Valeev, A. F. and Vreeswijk, P. M. and Walters, R. and Wozniak, P.},
	month = jul,
	year = {2017},
	pages = {46},
}

@article{blagorodnova_broad_2019,
	title = {The {Broad} {Absorption} {Line} {Tidal} {Disruption} {Event} {iPTF15af}: {Optical} and {Ultraviolet} {Evolution}},
	volume = {873},
	issn = {0004-637X, 1538-4357},
	shorttitle = {The {Broad} {Absorption} {Line} {Tidal} {Disruption} {Event} {iPTF15af}},
	url = {https://iopscience.iop.org/article/10.3847/1538-4357/ab04b0},
	doi = {10.3847/1538-4357/ab04b0},
	number = {1},
	urldate = {2025-10-21},
	journal = {The Astrophysical Journal},
	author = {Blagorodnova, N. and Cenko, S. B. and Kulkarni, S. R. and Arcavi, I. and Bloom, J. S. and Duggan, G. and Filippenko, A. V. and Fremling, C. and Horesh, A. and Hosseinzadeh, G. and Karamehmetoglu, E. and Levan, A. and Masci, F. J. and Nugent, P. E. and Pasham, D. R. and Veilleux, S. and Walters, R. and Yan, L. and Zheng, W.},
	month = mar,
	year = {2019},
	pages = {92},
}

@article{holoien_discovery_2019,
	title = {Discovery and {Early} {Evolution} of {ASASSN}-19bt, the {First} {TDE} {Detected} by {TESS}},
	volume = {883},
	issn = {0004-637X, 1538-4357},
	url = {https://iopscience.iop.org/article/10.3847/1538-4357/ab3c66},
	doi = {10.3847/1538-4357/ab3c66},
	number = {2},
	urldate = {2025-10-21},
	journal = {The Astrophysical Journal},
	author = {Holoien, Thomas W.-S. and Vallely, Patrick J. and Auchettl, Katie and Stanek, K. Z. and Kochanek, Christopher S. and French, K. Decker and Prieto, Jose L. and Shappee, Benjamin J. and Brown, Jonathan S. and Fausnaugh, Michael M. and Dong, Subo and Thompson, Todd A. and Bose, Subhash and Neustadt, Jack M. M. and Cacella, P. and Brimacombe, J. and Kendurkar, Malhar R. and Beaton, Rachael L. and Boutsia, Konstantina and Chomiuk, Laura and Connor, Thomas and Morrell, Nidia and Newman, Andrew B. and Rudie, Gwen C. and Shishkovksy, Laura and Strader, Jay},
	month = oct,
	year = {2019},
	pages = {111},
}

@article{arcavi_continuum_2014,
	title = {A {Continuum} of {H}- to {He}-{Rich} {Tidal} {Disruption} {Candidates} {With} a {Preference} for {E}+{A} {Galaxies}},
	copyright = {arXiv.org perpetual, non-exclusive license},
	url = {https://arxiv.org/abs/1405.1415},
	doi = {10.48550/ARXIV.1405.1415},
    journal = {The Astrophysical Journal},
	urldate = {2025-10-21},
	author = {Arcavi, Iair and Gal-Yam, Avishay and Sullivan, Mark and Pan, Yen-Chen and Cenko, S. Bradley and Horesh, Assaf and Ofek, Eran O. and De Cia, Annalisa and Yan, Lin and Yang, Chen-Wei and Howell, D. A. and Tal, David and Kulkarni, Shrinivas R. and Tendulkar, Shriharsh P. and Tang, Sumin and Xu, Dong and Sternberg, Assaf and Cohen, Judith G. and Bloom, Joshua S. and Nugent, Peter E. and Kasliwal, Mansi M. and Perley, Daniel A. and Quimby, Robert M. and Miller, Adam A. and Theissen, Christopher A. and Laher, Russ R.},
	year = {2014},
	note = {Publisher: arXiv
Version Number: 2},
}

@article{saxton_tidal_2012,
	title = {A tidal disruption-like {X}-ray flare from the quiescent galaxy {SDSS} {J120136}.02+300305.5},
	volume = {541},
	issn = {0004-6361, 1432-0746},
	url = {http://www.aanda.org/10.1051/0004-6361/201118367},
	doi = {10.1051/0004-6361/201118367},
	urldate = {2025-10-21},
	journal = {Astronomy \& Astrophysics},
	author = {Saxton, R. D. and Read, A. M. and Esquej, P. and Komossa, S. and Dougherty, S. and Rodriguez-Pascual, P. and Barrado, D.},
	month = may,
	year = {2012},
	pages = {A106},
}

@article{chornock_uv-bright_2013,
	title = {The {UV}-bright, {Slowly} {Declining} {Transient} {PS1}-11af as a {Partial} {Tidal} {Disruption} {Event}},
	copyright = {arXiv.org perpetual, non-exclusive license},
	url = {https://arxiv.org/abs/1309.3009},
	doi = {10.48550/ARXIV.1309.3009},
    journal = {The Astrophysical Journal},
	urldate = {2025-10-21},
	author = {Chornock, R. and Berger, E. and Gezari, S. and Zauderer, B. A. and Rest, A. and Chomiuk, L. and Kamble, A. and Soderberg, A. M. and Czekala, I. and Dittmann, J. and Drout, M. and Foley, R. J. and Fong, W. and Huber, M. E. and Kirshner, R. P. and Lawrence, A. and Lunnan, R. and Marion, G. H. and Narayan, G. and Riess, A. G. and Roth, K. C. and Sanders, N. E. and Scolnic, D. and Smartt, S. J. and Smith, K. and Stubbs, C. W. and Tonry, J. L. and Burgett, W. S. and Chambers, K. C. and Flewelling, H. and Hodapp, K. W. and Kaiser, N. and Magnier, E. A. and Martin, D. C. and Neill, J. D. and Price, P. A. and Wainscoat, R.},
	year = {2013},
	note = {Publisher: arXiv
Version Number: 1},
}

@article{bloom_possible_2011,
	title = {A {Possible} {Relativistic} {Jetted} {Outburst} from a {Massive} {Black} {Hole} {Fed} by a {Tidally} {Disrupted} {Star}},
	volume = {333},
	issn = {0036-8075, 1095-9203},
	url = {https://www.science.org/doi/10.1126/science.1207150},
	doi = {10.1126/science.1207150},
	language = {en},
	number = {6039},
	urldate = {2025-10-21},
	journal = {Science},
	author = {Bloom, Joshua S. and Giannios, Dimitrios and Metzger, Brian D. and Cenko, S. Bradley and Perley, Daniel A. and Butler, Nathaniel R. and Tanvir, Nial R. and Levan, Andrew J. and O' Brien, Paul T. and Strubbe, Linda E. and De Colle, Fabio and Ramirez-Ruiz, Enrico and Lee, William H. and Nayakshin, Sergei and Quataert, Eliot and King, Andrew R. and Cucchiara, Antonino and Guillochon, James and Bower, Geoffrey C. and Fruchter, Andrew S. and Morgan, Adam N. and Van Der Horst, Alexander J.},
	month = jul,
	year = {2011},
	pages = {203--206},
}

@article{holoien_asassn-14ae_2014,
	title = {{ASASSN}-14ae: {A} {Tidal} {Disruption} {Event} at 200 {Mpc}},
	copyright = {arXiv.org perpetual, non-exclusive license},
	shorttitle = {{ASASSN}-14ae},
	url = {https://arxiv.org/abs/1405.1417},
	doi = {10.48550/ARXIV.1405.1417},
    journal = {Monthly Notices of the Royal Astronomical Society},
	urldate = {2025-10-21},
	author = {Holoien, Thomas W. -S. and Prieto, J. L. and Bersier, D. and Kochanek, C. S. and Stanek, K. Z. and Shappee, B. J. and Grupe, D. and Basu, U. and Beacom, J. F. and Brimacombe, J. and Brown, J. S. and Davis, A. B. and Jencson, J. and Pojmanski, G. and Szczygiel, D. M.},
	year = {2014},
	note = {Publisher: arXiv
Version Number: 2},
}

@article{holoien_six_2016,
	title = {Six months of multiwavelength follow-up of the tidal disruption candidate {ASASSN}-14li and implied {TDE} rates from {ASAS}-{SN}},
	volume = {455},
	issn = {0035-8711, 1365-2966},
	url = {https://academic.oup.com/mnras/article-lookup/doi/10.1093/mnras/stv2486},
	doi = {10.1093/mnras/stv2486},
	language = {en},
	number = {3},
	urldate = {2025-10-21},
	journal = {Monthly Notices of the Royal Astronomical Society},
	author = {Holoien, T. W.-S. and Kochanek, C. S. and Prieto, J. L. and Stanek, K. Z. and Dong, Subo and Shappee, B. J. and Grupe, D. and Brown, J. S. and Basu, U. and Beacom, J. F. and Bersier, D. and Brimacombe, J. and Danilet, A. B. and Falco, E. and Guo, Z. and Jose, J. and Herczeg, G. J. and Long, F. and Pojmanski, G. and Simonian, G. V. and Szczygieł, D. M. and Thompson, T. A. and Thorstensen, J. R. and Wagner, R. M. and Woźniak, P. R.},
	month = jan,
	year = {2016},
	pages = {2918--2935},
}

@article{dong_asassn-15lh_2016,
	title = {{ASASSN}-15lh: {A} highly super-luminous supernova},
	volume = {351},
	issn = {0036-8075, 1095-9203},
	shorttitle = {{ASASSN}-15lh},
	url = {https://www.science.org/doi/10.1126/science.aac9613},
	doi = {10.1126/science.aac9613},
	language = {en},
	number = {6270},
	urldate = {2025-10-21},
	journal = {Science},
	author = {Dong, Subo and Shappee, B. J. and Prieto, J. L. and Jha, S. W. and Stanek, K. Z. and Holoien, T. W.-S. and Kochanek, C. S. and Thompson, T. A. and Morrell, N. and Thompson, I. B. and Basu, U. and Beacom, J. F. and Bersier, D. and Brimacombe, J. and Brown, J. S. and Bufano, F. and Chen, Ping and Conseil, E. and Danilet, A. B. and Falco, E. and Grupe, D. and Kiyota, S. and Masi, G. and Nicholls, B. and Olivares E., F. and Pignata, G. and Pojmanski, G. and Simonian, G. V. and Szczygiel, D. M. and Woźniak, P. R.},
	month = jan,
	year = {2016},
	pages = {257--260},
}

@article{kajava_rapid_2020,
	title = {Rapid late-time {X}-ray brightening of the tidal disruption event {OGLE16aaa}},
	volume = {639},
	copyright = {https://www.edpsciences.org/en/authors/copyright-and-licensing},
	issn = {0004-6361, 1432-0746},
	url = {https://www.aanda.org/10.1051/0004-6361/202038165},
	doi = {10.1051/0004-6361/202038165},
	urldate = {2025-10-21},
	journal = {Astronomy \& Astrophysics},
	author = {Kajava, Jari J. E. and Giustini, Margherita and Saxton, Richard D. and Miniutti, Giovanni},
	month = jul,
	year = {2020},
	pages = {A100},
}

@article{hung_revisiting_2017,
	title = {Revisiting {Optical} {Tidal} {Disruption} {Events} with {iPTF16axa}},
	volume = {842},
	issn = {0004-637X, 1538-4357},
	url = {https://iopscience.iop.org/article/10.3847/1538-4357/aa7337},
	doi = {10.3847/1538-4357/aa7337},
	number = {1},
	urldate = {2025-10-21},
	journal = {The Astrophysical Journal},
	author = {Hung, T. and Gezari, S. and Blagorodnova, N. and Roth, N. and Cenko, S. B. and Kulkarni, S. R. and Horesh, A. and Arcavi, I. and McCully, C. and Yan, Lin and Lunnan, R. and Fremling, C. and Cao, Y. and Nugent, P. E. and Wozniak, P.},
	month = jun,
	year = {2017},
	pages = {29},
}

@article{johansson_ztf_2023,
	title = {{ZTF} {Transient} {Classification} {Report} for 2023-05-12},
	volume = {2023-1075},
	url = {https://ui.adsabs.harvard.edu/abs/2023TNSCR1075....1J},
	urldate = {2025-10-21},
	journal = {Transient Name Server Classification Report},
	author = {Johansson, J. and Meynardie, W. and Chu, M. and Fremling, C.},
	month = may,
	year = {2023},
	note = {ADS Bibcode: 2023TNSCR1075....1J},
	pages = {1},
}

@article{van_velzen_first_2019,
	title = {The {First} {Tidal} {Disruption} {Flare} in {ZTF}: {From} {Photometric} {Selection} to {Multi}-wavelength {Characterization}},
	volume = {872},
	issn = {0004-637X, 1538-4357},
	shorttitle = {The {First} {Tidal} {Disruption} {Flare} in {ZTF}},
	url = {https://iopscience.iop.org/article/10.3847/1538-4357/aafe0c},
	doi = {10.3847/1538-4357/aafe0c},
	number = {2},
	urldate = {2025-10-21},
	journal = {The Astrophysical Journal},
	author = {van Velzen, Sjoert and Gezari, Suvi and Cenko, S. Bradley and Kara, Erin and Miller-Jones, James C. A. and Hung, Tiara and Bright, Joe and Roth, Nathaniel and Blagorodnova, Nadejda and Huppenkothen, Daniela and Yan, Lin and Ofek, Eran and Sollerman, Jesper and Frederick, Sara and Ward, Charlotte and Graham, Matthew J. and Fender, Rob and Kasliwal, Mansi M. and Canella, Chris and Stein, Robert and Giomi, Matteo and Brinnel, Valery and Santen, Jakob Van and Nordin, Jakob and Bellm, Eric C. and Dekany, Richard and Fremling, Christoffer and Golkhou, V. Zach and Kupfer, Thomas and Kulkarni, Shrinivas R. and Laher, Russ R. and Mahabal, Ashish and Masci, Frank J. and Miller, Adam A. and Neill, James D. and Riddle, Reed and Rigault, Mickael and Rusholme, Ben and Soumagnac, Maayane T. and Tachibana 優太, Yutaro 朗橘},
	month = feb,
	year = {2019},
	pages = {198},
}

@article{mummery_optical_2025,
	title = {The optical, {UV}-plateau, and {X}-ray tidal disruption event luminosity functions reproduced from first principles},
	volume = {541},
	copyright = {https://creativecommons.org/licenses/by/4.0/},
	issn = {0035-8711, 1365-2966},
	url = {https://academic.oup.com/mnras/article/541/1/429/8178511},
	doi = {10.1093/mnras/staf938},
	language = {en},
	number = {1},
	urldate = {2025-10-21},
	journal = {Monthly Notices of the Royal Astronomical Society},
	author = {Mummery, Andrew and van Velzen, Sjoert},
	month = jun,
	year = {2025},
	pages = {429--445},
}

@article{leloudas_spectral_2019,
	title = {The {Spectral} {Evolution} of {AT} 2018dyb and the {Presence} of {Metal} {Lines} in {Tidal} {Disruption} {Events}},
	volume = {887},
	issn = {0004-637X, 1538-4357},
	url = {https://iopscience.iop.org/article/10.3847/1538-4357/ab5792},
	doi = {10.3847/1538-4357/ab5792},
	number = {2},
	urldate = {2025-10-21},
	journal = {The Astrophysical Journal},
	author = {Leloudas, Giorgos and Dai, Lixin and Arcavi, Iair and Vreeswijk, Paul M. and Mockler, Brenna and Roy, Rupak and Malesani, Daniele B. and Schulze, Steve and Wevers, Thomas and Fraser, Morgan and Ramirez-Ruiz, Enrico and Auchettl, Katie and Burke, Jamison and Cannizzaro, Giacomo and Charalampopoulos, Panos and Chen, Ting-Wan and Cikota, Aleksandar and Della Valle, Massimo and Galbany, Lluis and Gromadzki, Mariusz and Heintz, Kasper E. and Hiramatsu, Daichi and Jonker, Peter G. and Kostrzewa-Rutkowska, Zuzanna and Maguire, Kate and Mandel, Ilya and Nicholl, Matt and Onori, Francesca and Roth, Nathaniel and Smartt, Stephen J. and Wyrzykowski, Lukasz and Young, Dave R.},
	month = dec,
	year = {2019},
	pages = {218},
}
